\documentclass[useAMS,usenatbib,a4paper, fleqn]{mnras}
\usepackage{ae,aecompl}
\title[Cluster formation in galactic flows]
{The formation of massive stellar clusters in converging galactic flows with photoionisation
}
\author[Dobbs]
{C. L. Dobbs\thanks{E-mail:
C.L.Dobbs@exeter.c.uk}$^{1}$, T. J. R. Bending$^{1}$, A. R. Pettitt$^{2,3}$, M. R. Bate$^{1}$\\
$^{1}$School of Physics and Astronomy, University of Exeter, Stocker Road, Exeter, EX4 4QL, UK\\
$^{2}$Department of Physics, Faculty of Science, Hokkaido University, Sapporo 060-0810, Japan\\
$^{3}$Department of Physics and Astronomy, California State University, Sacramento, 6000 J Street, Sacramento, CA 95819-6041, USA\\
}
\usepackage{amssymb}
\usepackage{amsmath}
\usepackage[pdftex]{graphicx}
\usepackage{epsfig}
\usepackage{multirow} 
\def\aap{{ A\&A}}

\def\apjl{{ApJL}}

\def\apjs{{ApJS}}

\begin{document}
\label{firstpage}
\date{\today}

\pagerange{\pageref{firstpage}--\pageref{lastpage}} \pubyear{2021}

\maketitle

\begin{abstract}
We have performed simulations of cluster formation along two regions of a spiral arm taken from a global Milky Way simulation, including photoionising feedback. One region is characterised by strongly converging flows, the other represents a more typical spiral arm region. We find that more massive clusters are able to form on shorter timescales for the region with strongly converging flows. Mergers between clusters are frequent in the case of the strongly converging flows and enable the formation of massive clusters. We compare equivalent clusters formed in simulations with and without ionisation. Photoionisation does not prevent massive cluster formation, but can be seen to limit the masses of the clusters. On average the mass is reduced by around 20\%, but we see a large spread from ionisation having minimal difference to leading to a 50\% reduction in mass. Photoionisation is also able to clear out the gas in the vicinity of the clusters on Myr timescales,  which can produce clusters with larger radii that are surrounded by more massive stellar halos. We find that the ionising feedback has more impact in our second region which is less dense and has less strongly converging flows.
\end{abstract} 

\begin{keywords}
stars: formation, ISM: clouds, galaxies: star formation, galaxies: star clusters: general
\end{keywords}

\section{Introduction}
Understanding how stellar clusters, and in particular how different types of cluster form, is one of the fundamental problems in star formation. Much numerical work has investigated cluster formation, following e.g. \citet{Bate2003}, \citet{Klessen2000} and \citep{McKee2003}, but most have considered cluster formation in a turbulent box or sphere. As such these works simply form one cluster with properties strongly dependent on the initial conditions. Cluster formation has been investigated in the context of dwarf galaxies \citep{Hu2016,Emerick2018,Lahen2019}, where the gas mass is sufficiently low that high resolution in the gas and stars can realistically be achieved, but less so in typical Milky Way type environments.

In previous work, we have investigated massive cluster formation in colliding flows \citep{Dobbs2020,Liow2020,Dobbs2021}. We found that cluster formation is enhanced by collisions, but that the velocity needs to be quite high, around 20 km s$^{-1}$ (see also \citealt{Maeda2021}). At such velocities, more massive clusters form on shorter timescales, compared to isolated clouds. However these studies lack a more realistic galactic context. Simulations by Rieder et al., submitted, investigate cluster formation in simulations with fixed spiral arms of different strengths, which mimic different convergence of the gas flows in spiral arms. \citet{Smilgys2017} also modelled clusters along a section of spiral arm, finding that more massive clusters form by hierarchical merging.
In reality, spiral arms will not necessarily be fixed. \citet{Pettitt2015} modelled the Milky Way and actually found that a better fit to the observed spiral arms was obtained when the spiral arms were dynamically evolving compared to fixed \citep{Pettitt2014}. Considering more complex dynamics potentially allows more variation in the flows, including even for example spiral arms merging. 

Another question is the effect of feedback on cluster formation and evolution. Again, detailed studies of feedback have largely been applied to studies of individual clouds, and dwarf galaxies. Neither the work of Rieder et al. or \citet{Smilgys2017} include feedback. For short timescales, similar to those to form young massive clusters, only immediate feedback processes such as ionisation, winds and radiation pressure are likely to be important, since supernovae will not have had time to occur.

Simulations of isolated clouds show that ionisation can strongly affect the morphology of molecular clouds, reducing the gas mass by 10s of per cent, although this effect is smaller for higher mass or density clouds \citep{Dale2012,Dale2017,Colin2013,Gavagnin2017,Ali2019,Kim2021b}. \citet{Geen2018} do not appear to find that strong feedback has a comparable effect on the stellar cluster. \citet{Geen2017} find that for the densest clouds, the star formation efficiency can still be very high (10s per cent), whilst at sufficiently high densities, the star formation efficiency can tend to 100 per cent \citep{Grudic2018}. In contrast to some of these studies which consider decaying turbulence, \citet{Sartorio2021} find that if the turbulence is continuously driven, the effects from ionising feedback are reduced.

On galaxy scales, simulations of dwarf galaxies find that ionisation has a strong effect on clearing out gas around young stars \citep{Semenov2021} supporting the idea that ionisation dominates cloud lifetimes \citep{Kruijssen2019,Chevance2020}. Surveys of clusters in nearby galaxies also find that ionisation also appears to disperse gas and dust in the vicinity of clusters on very short, $\sim$ Myr timescales \citep{Kawamura2009,Holyhead2015,Grasha2019,Hannon2019,Barnes2020,Messa2021,Kim2021}. The density of the cloud, or the potential well of the gas in which the cluster forms may however also play a role in how easy it is for gas to be affected by ionisation. For example, star clusters may remain embedded longer when formed in clouds with deeper gravitational potentials \citep{Dinnbier2020,Hajime2021}.

\citet{Grudic2020} and \citet{Geen2020} find that photoionisation is particularly important on GMC scales, whilst supernovae may have a greater impact on larger scales \citep{Semenov2021}. 
The effect of winds on star formation and the surrounding gas is generally found to be less than ionisation \citep{Dale2013}. Supernovae may also be less effective, since once earlier forms of feedback occur, supernovae simply fill the ionised regions with hot gas \citep{Walch2015} (see also Bending et al. in prep.). Hot gas can also escape through channels into the wider ISM rather than significantly effecting the colder, denser medium \citep{Lucas2020}.
Radiation pressure appears to dominate in the earliest stages after massive stars start to undergo feedback, on relatively small scales \citep{Barnes2020,Ali2021}, and may also be more important in starburst regimes \citep{Krumholz2012,Geen2020}. \citet{Sales2014} find that generally photoionisation dominates radiation pressure, although \citet{Tsang2018} find that radiation pressure appears to reduce the star formation efficiency in massive clusters by around one third compared to having no feedback, but radiation pressure does not suppress continual growth of the clusters.

To date, most studies involve following clusters formed in single clouds, or lack the resolution to follow clusters.
As such they cannot follow clusters formed in a realistic galactic environment, or study the effects of feedback in a larger scale context.
In this paper we look at cluster formation along two sections of spiral arm including photoionisation, but not other forms of feedback. We choose regions which have contrasting dynamics - one is from a region which is strongly converging, and the other is a region which is only showing moderate gas convergence. Both regions are from a galactic scale model of the Milky Way, where the spiral arms form self consistently. As such we test the impact of the galactic scale environment on the clusters which form, and also the impact of photoionisation on clusters which arise self consistently from realistic galactic conditions. 

In Section~\ref{method} we describe the extraction of our initial conditions and computational method. We describe our results in Sections~\ref{Set1results} and \ref{Set2results}, and in particular we show cluster evolution and properties in Sections~\ref{Clusters} and \ref{properties}. Finally in Section~\ref{conclusions} we discuss our results and present our conclusions.

\section{Method}\label{method}
\subsection{Initial conditions}
\subsubsection{Set 1 simulations}\label{Set1initial}
We perform simulations of sections of galaxy simulations using the Smoothed Particle Hydrodynamics code sphNG \citep{Benz1990,Bate1995,Price2007}.
We discuss our initial conditions here, and then in Section 2.2 we discuss the physics we include in our calculations. We discuss the initial conditions first to give more context to the relevant size scales, and description of the physical methods.

We take our galaxy sections from the simulation of the Milky Way labelled `Normal disc' by \citet{Pettitt2015}, which includes a live stellar component, and a NFW dark matter halo. These past simulations include heating and cooling, H$_2$ and CO formation, though they do not include gas self gravity, star formation or stellar feedback. We use the model which produces a spiral arm pattern which is in extremely good agreement with the Milky Way in the mid to outer disc. The initial galaxy model contains 1100008 star particles, and 3 million gas particles, which corresponds to a gas mass per particle of 2666 M$_{\odot}$. The initial gas density profile was set up to follow the stellar profile. Star particles have adaptive softening according to \citet{Price2007}. As we discuss in Section 2.2, all our simulations here use the same NFW dark matter halo, stellar disc, heating and cooling as the original simulations, but now we include gas self gravity, star formation and photoionising feedback.

We perform simulations focusing on two different regions from the global Milky Way model, Region 1 and Region 2. Both of these regions are along spiral arms, but we choose regions with differing gas flows.  Region 1 is strongly converging whereas Region 2 is less strongly converging, although both regions are along the same spiral arm.
 \begin{figure*}
\centerline{\includegraphics[scale=0.7]{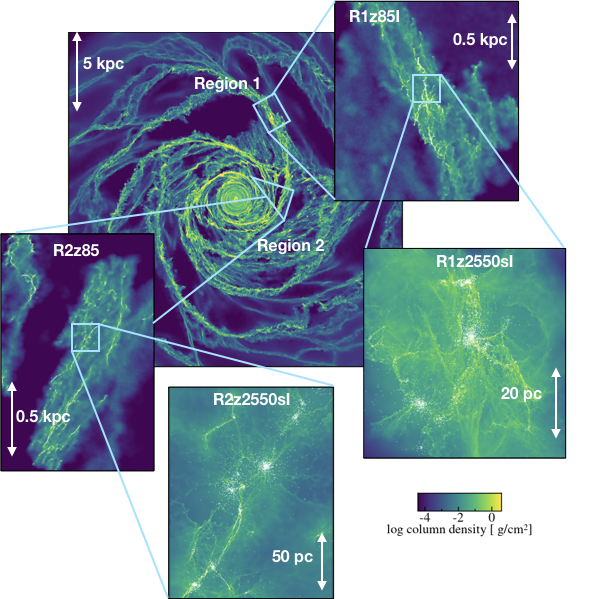}} 
\caption{This figure shows a panel from the whole galaxy simulation \citep{Pettitt2015} which we used for our initial conditions (background), panels from Set 1 Region 1 and Region 2 simulations indicating where they lie in the galaxy, and panels from two of the Set 2 resimulations indicating where they are located.}   
\label{fig:collage}
\end{figure*}

We show in Figure~\ref{fig:collage} the column density plot of the galaxy scale simulation, and the regions that we resimulate. We also show a list of the simulations carried out in Table~1. We initially ran calculations with the same surface density as the simulations in \citet{Pettitt2015}. However after finding quite high star formation rates, we noticed that although matching at larger radii, the surface density was a factor of around 2 too high at inner radii compared with the Milky Way. So we halved the gas mass compared to the original calculation. 
In the original simulations, the gas distribution will be predominantly determined by the gravity from the stars and dark matter, and sound speed of the gas, rather than the gas mass itself, so reducing the gas mass is still consistent with the other initial gas properties, and in better agreement with the Milky Way.
For the Set 1 simulations (see Table~1), we set up the initial conditions for each region as follows. We select the gas particles at a timeframe of 207 Myr, and trace these particles back to a time of 198 Myr. We split the particles by a factor of 85, to a mass of 15 M$_{\odot}$ (using the method described in \citealt{Bending2020}). For Region 1, the initial conditions are a sheared region of approximately 1-2 kpc squared in the $xy$ plane, and a vertical distribution up to $\sim$400 pc. After particle splitting, this simulation has 5391210 gas particles, and gas mass of 8.46 $\times 10^7$ M$_{\odot}$. For Region 2, the initial conditions are an approximately triangular region of base 1 kpc, height 2 kpc in the $x-y$ plane, and a vertical thickness of 200-400 pc. This simulation has 5974500 gas particles, and a gas mass of 7.65 $\times 10^7$ M$_{\odot}$. Surrounding gas is not included, but both simulations contain all 1100008 star particles present in the original simulation, which exhibit the global spiral pattern of the disc. Gravity softening of the stars is coupled to the gas mass \citep{Price2007} so the gravity resolution is adapted to the new resolution in the hydrodynamics.

In our Set 1 simulations we split the gas particles by a factor of 85 compared to the original galaxy scale simulations. However the mass of gas particles was still quite large in this instance (15 M$_{\odot}$). As described in Section~{\ref{Details}, we include sink particles to model star formation, and since sink particles typically form from a few 10s of gas particles, the sink particles are also quite massive. Hence we zoom in further on subregions of our Set 1 simulations, splitting the particles by a further factor of 30, and these further zoom in calculations we refer to as Set 2.

In our zoom in simulations we do not model the surrounding gas, so naturally information is lost in resimulations which might affect gas at the edges of our regions. However we focus on regions away from the edges of the simulations, so for example the Set 2 simulations are taken from the middle of the Set 1 simulations, and the clusters we focus on in Set 2 are not near the edge, so this avoids the boundary conditions having a significant effect on the results. 

From herein we use the time of 198 Myr, at which we set up the initial conditions for the Set 1 simulations as $t=0$\,Myr, and all subsequent times are listed relative to this time.

\subsubsection{Set 2 simulations}
To better study cluster formation and the effects of ionisation, we resimulate sections of Regions 1 and 2. Splitting each particle by a factor of 30, this obtains a mass per particle of 0.5~M$_{\odot}$. We call these Set 2 simulations, and denote them `z2550' to indicate a factor of 2550 times higher resolution than the original simulation. We indicate the sections of Regions 1 and 2 which we resimulate in Figure~\ref{fig:collage}.

To set up the Set 2 simulations, we select particles at a time of 1.2 Myr from models R1z85Is and R2z85s, and trace them back to an earlier time of 0.6 Myr. We then selected a box at the 0.6 Myr time which contained all the particles which had been traced back. This also meant that other neighbouring particles which lay in the box were selected as well.
We originally tried selecting a region towards the end of the Set 1 simulations, and tracing the gas back to resimulate.  However this proved to be unsuitable as 0.6 Myr prior to the end of the Set 1 simulations, at least in the R1z85I simulation, massive clusters had already formed. If tracing back over longer timescales ($>1$ Myr), the locus of particles was large, and thus a box which contained all these particles was so large that the resimulation would have required 10's of millions of particles. 
 
 We also added a small ($\sim$ 1 km s$^{-1}$) velocity dispersion to the particles at the start of the Set 2 simulations, to mimic small scale unresolved velocities in the Set 1 simulations, however this did not greatly effect the simulation, and after trying a few different values, the simulations were the same after a short time unless very large ($\mathcal{O}(50)$km s$^{-1}$ or so) dispersions were used. 

For the Set 2 Region 1 simulations (R1z2550s and R1z2550Is), we chose a region at around $x=2.5$ kpc, $y=4.75$ kpc in Figure~2, where we clearly see clusters forming.  The initial size of our resimulated region is approximately 70 pc $\times$ 70 pc, and contains $2.75\times10^6$ M$_{\odot}$ of gas and of stars (situated in sink particles). We also split the sink particles by a factor of 30, using the accretion radius (0.2 pc) instead of the smoothing length to distribute the new particles. For those sink particles which contain ionising sources, we assign the ionising flux to the sink particle at the location of the original sink before it is split, and set the ionisation to zero for the other particles. For Region 2, there are relatively few sink particles at 0.6 Myr. The region we choose, which lies in the spiral arm section of Region 2, does not contain any sink particles. This is not unrepresentative of the simulation at this point, and avoids the complication of splitting the sink particles. This section is approximately 160 $\times$ 160 pc, and contains $2.5\times10^6$ M$_{\odot}$ of gas. For both the Set 2 Region 1 simulations (R1z2550s and R1z2550Is), and Set 2 Region 2 simulations (R2z2550s and R2z2550Is), we run models with and without ionisation.
\begin{table*}
\begin{tabular}{c | c|c|c|c | c}
 \hline 
 \hline 
Name & Seed sim. & Region & Particle mass & Ionisation? & Sinks \\
& & & (M$_{\odot}$) & & \\
 \hline 
 \textbf{Set 1} & & & & \\
 \hline 
R1z85 & Disc & 1& 15 & N & Normal \\
R1z85I  & Disc & 1& 15 & Y & Normal \\
R2z85  & Disc & 2& 15 & N & Normal \\
R2z85I  & Disc & 2& 15 & Y & Normal \\
R1z85s & Disc & 1& 15 & N & Small \\
R1z85sI  & Disc & 1& 15 & Y & Small \\
R2z85s  & Disc & 2& 15 & N & Small\\
 \hline 
 \textbf{Set 2} & & & & \\
 \hline 
R1z2550s &   R1z85s  &1 & 0.5 & N & Small \\
R1z2550sI  & R1z85s &1 & 0.5 & Y & Small \\
R2z2550s  &  R2z85s &2 & 0.5 & N & Small \\
R2z2550sI  & R2z85s &2 & 0.5 & Y & Small \\
\hline
 \hline 
\end{tabular}
\caption{Table listing the simulations carried out, whether they include ionisation and the sink particle parameters (see text). For Set 1 the seed simualtion is the global galaxy simulation, whereas for Set 2, the seed simulation is one of the Set 1 simulations. The Set 2 simulations are of smaller regions and higher resolution compared to Set 1.}\label{tab:cloudtable}
\end{table*}

\subsubsection{Convergence of gas}
We can make a simple estimate of the mass which could converge in our regions following \citet{Dobbs2020}.  We estimate the mass which can converge in this region over a given timescale according to 
\begin{equation}
M=v \rho A t
\end{equation}
where $v$ is the relative velocity between the converging flows, $\rho$ is their density, $A$ is the cross sectional area in the arm, and $t$ is some timescale. Taking a distance scale of 100 pc, which is similar to the size scale of our Set 2 simulations, 
we find that for Region 1, the relative velocity is around 20 km s$^{-1}$. Taking a density of 100 cm$^{-3}$ we would expect a mass well in excess of $10^5$ M$_{\odot}$ to be able to converge over 100 pc in Region 1 on a timescale of 1 Myr. For Region 2, the relative velocities of the gas are generally less than 10 km~s$^{-1}$. We show the divergence of the velocity field in Figure~\ref{fig:convergingflows}, with Regions 1 and 2 overlaid. The divergence is calculated on a particle by particle basis over a smoothing length. This likewise shows that the convergence is higher for Region 1 compared to Region 2.
\begin{figure}
\centerline{\includegraphics[scale=0.41]{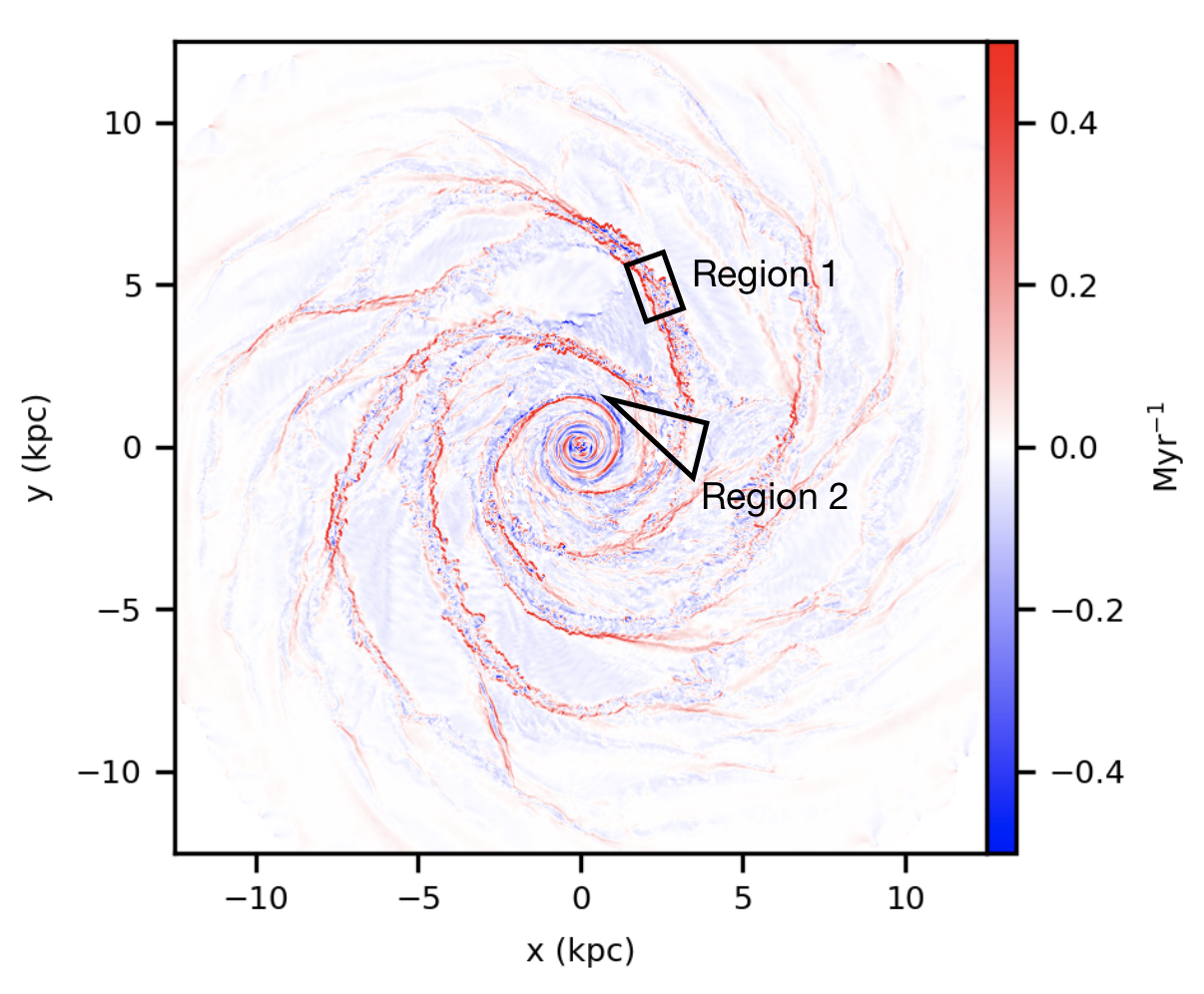}} \caption{The divergence of the velocity field is shown for the galaxy model from \citet{Pettitt2015}. The divergence is calculated on a particle by particle basis. Region 1 has a more negative divergence indicating more strongly converging flows, compared to Region 2.}  
\label{fig:convergingflows}
\end{figure}

As such we would predict that Region 1 is a prime site for forming massive clusters on short timescales. The strong convergence generally along this section of spiral arm appears to be linked to the merger of this spiral arm with another arm (see Figure~\ref{fig:collage}). The effect of this merger can be seen from the shape of the spiral arm and the kink just above Region 1, whilst the merger of two spiral arms may mean gas convergence is not just limited to across (perpendicular to) the spiral arm, but along the spiral arm as well. A similar phenomenon is seen in the next innermost spiral arm as well.

\subsection{Details of Simulations}\label{Details}

In all our simulations, we model the full halo and stellar components of the galaxy. We model the gas just in our selected zoomed in regions. These are the same as the global galaxy model. We use the same analytic potential for the dark matter halo as \citet{Pettitt2015}. Stars in the disc and bulge present at the simulation start are modelled as live components, represented as $N$-body particles that interact only gravitationally with other particles (acting as an ``old" stellar population). For the gas, we apply the same heating and cooling prescription, and the same chemistry prescription for H$_2$ and CO formation as the original model. Most of our gas is molecular (we use the same simple fixed length scale (35 pc) for the self shielding of the molecular hydrogen, although the high densities suggest that the gas would likely largely be molecular anyway). In addition to the physics used in the galaxy scale models, we now add self gravity, star formation and photoionising feedback.

In all our simulations, sink particles are formed once a given density threshold of 1000 cm$^{-3}$ is exceeded, but similar to \citet{Bending2020}, gas can exceed this density and not meet the criteria for sink formation (see \citealt{Bending2020} and \citealt{Bate1995}). Again, like \citet{Bending2020} we impose a density of $10^5$ cm$^{-3}$ above which sink formation occurs whether or not the criteria are met. Sink particles represent stellar populations, and can contain 0, 1 or multiple ionising stars. Sink particles can undergo mergers, which occurs automatically if sink particles come within a certain distance of each other. In Table 1 we list the sinks as either `Standard' or `Small'. We initially used an accretion radius of 0.5 pc, such that matter within this radius is accreted subject to checks (namely that the gas is convergent on the sink particle), and a merger radius of 0.01 pc. We list this as the `Standard' scheme in Table~1. However we found that the mass of the sink particles in the simulations could be very massive ($10^5$ M$_{\odot}$ although more typically the masses are $\sim 10^3$ M$_{\odot}$). So we also reran these simulations with an accretion radius of 0.2 pc and a merger radius of 0.001 pc, such that the largest sink particles are around a few $10^4$~M$_{\odot}$. In particular we use this sink prescription to produce the starting conditions for our second set of resimulations (Set 2).
Changing the sink parameters mainly means that there are more sink particles of lower mass. In the Set 2 simulations, we decreased the accretion radius further to 0.1 pc, but keep the merger radius the same. In all simulations (including the resimulations), the minimum temperature is 100 K, in order to resolve the Jeans mass in the simulations.
\begin{figure*}
\centerline{
\includegraphics[scale=0.54]{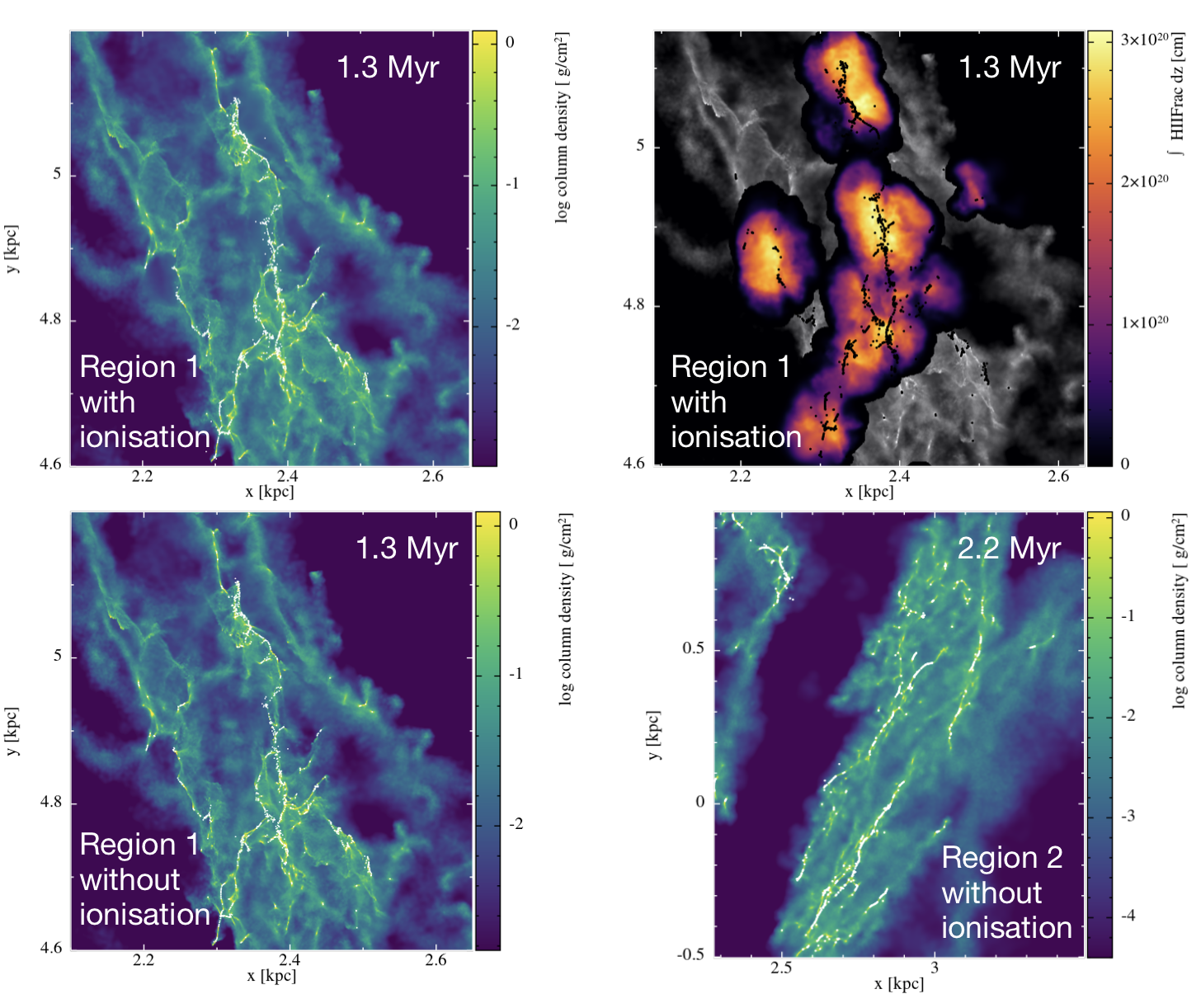}}
\caption{Column density plots are shown for the Set 1 simulations (R1z85, R1z85I and R2z85). The panels show Region 1 with ionisation (top row, right with \ion{H}{II} fraction overplotted on density), Region 1 without ionisation (lower left) and Region 2 (lower right). Sink particles are overplotted in white. Ionisation makes minimal difference at these scales. For Region 2, the sinks are mostly located in filamentary features.}  
\label{fig:set1}
\end{figure*}

For the photoionising feedback, we follow the scheme used in \citet{Bending2020}. To recap briefly on our previous work, we evolve the \ion{H}{II} fraction 
using SPH interpolation to integrate the contribution over all SPH particles whose smoothing lengths overlap with the line of sight between a given SPH particle and an ionising source.  
The ionisation fraction is evolved as
\begin{equation}
\begin{aligned}
\frac{{\rm d}H_{\rm II}}{{\rm d}t} = \frac{h^2}{r^2} \left( \frac{Q_{\rm H}}{4 \pi} \right. &  -  \int^{r}_{0} r^{\prime 2} n(r^\prime)^2 \alpha_{\rm B} {\rm d}r^\prime  \\ 
& -  \left. \frac{1}{\delta t} \int^{r}_{0} r^{\prime 2} n(r^\prime) [1-H_{\rm II}(r^\prime)] {\rm d}r^\prime  \right),
\label{eq:evol_ionisation}
\end{aligned}
\end{equation}
 where $h$ is the smoothing length, $Q_{\rm H}$ is the ionising flux, $n$ is the number density, $\alpha_B$ is the recombination efficiency (here $2.7\times 10^{-13}$ cm$^3$ s$^{-1}$), $\delta t$ is the time interval and \ion{H}{II} is the ionisation fraction of the gas (from 0 to 1). We heat gas which is ionised to $10^4$ K. To allow the simulations to be computationally feasible, we only integrate out to a maximum distance from the ionising sources. Here we only consider lines of sight for which $r<100$ pc. Stars are assigned massive stars via a sampling method tuned to reproduce a Kroupa IMF \citep{Kroupa2001}. The IMF samples masses between 0.01 and 107.5 M$_{\odot}$. Photoionisation is modelled for stars with masses $>18$ M$_{\odot}$. Typically sinks are of order 100 or a few 100's M$_{\odot}$, and thus well sample the IMF, but they do not all contain stars above masses of 18 M$_{\odot}$. Similarly to the SR\_50\% model in \citet{Bending2020} we choose an efficiency of 50\%, which means that half the mass in sinks is used to calculate the ionisation fluxes.

We made two small changes to the method used in \citet{Bending2020}. Firstly we changed 
the sampling slightly.
In the simulations presented in \citet{Bending2020}, new ionising stars were assigned to the sink particle with the highest non-ionising star mass. For the simulations presented here, we found a wider range in sink masses, which led to a disproportionately large number of ionising stars in the most massive sinks. Instead we assign the ionising star to the sink particle with the highest total non-star (i.e. unassigned) mass.
Secondly in the Set 2 simulations we changed the maximum line of sight from the 100 pc used by \citet{Bending2020} to approximately half the simulation sizescale. We checked this approach from the size of ionisation regions in our initial runs (which were relatively quicker) before we changed the sampling method, and also ran for a short time with different maximum truncation distances to check there was no significant difference. We also include a record of when stars formed will undergo supernovae in our simulations (Bending et al., in prep), however we do not run our simulations long enough for these stars to undergo supernovae.

\begin{figure}
\centerline{\includegraphics[scale=0.42]{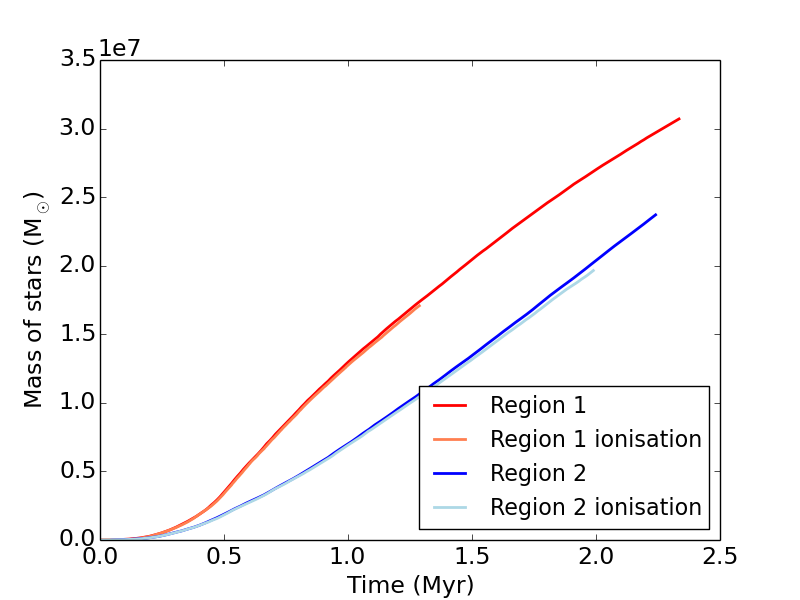}}
\centerline{\includegraphics[scale=0.42]{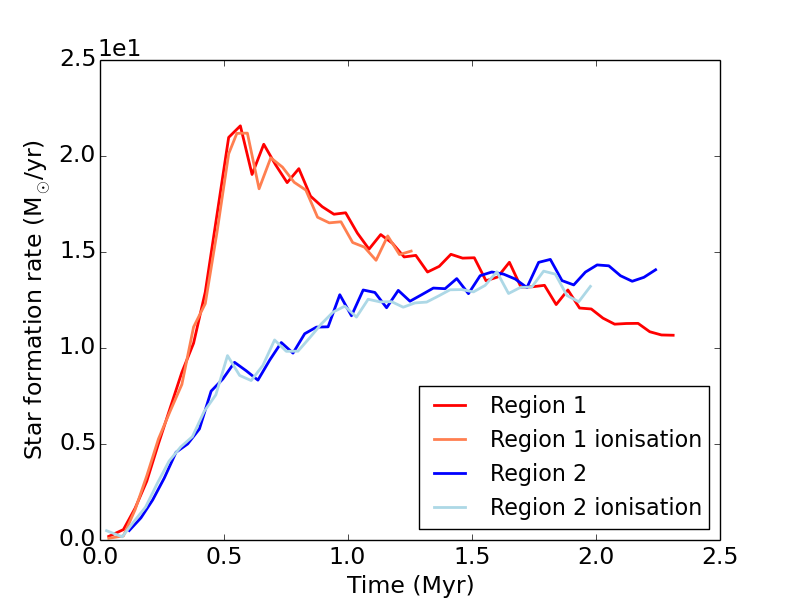}}
\caption{The mass of stars formed (top) and star formation rate
(lower) are shown versus for the Set 1 simulations of Region 1 and 2 without ionisation (R1z85, R1z85I, R2z85 and R2z85I). Region 2 forms fewer stars, and at least at early times has a significantly lower star formation rate. This is expected given the region lower velocity converging flows.}
\label{fig:sflarge}
\end{figure}

\section{Results for Set 1}\label{Set1results}  
In Figure~\ref{fig:set1} we show column density maps from three of our Set 1 simulations (R1z85, R1z85I and R2z85), where we have split the particles once and have a mass per particle of 15 M$_{\odot}$. On the left hand panels, and the top right panel, we show Region 1, where there are stronger converging flows. For both Regions 1 and 2, the sink particles form along filamentary structures, but there is more of a central area in Region 1 where there are multiple filaments merging together. In Region 2 (lower right panel), the sinks occur along more or less isolated filamentary structures. On these scales the presence of ionising feedback makes very little difference to the large scale structure, and it is difficult to see any differences between Region 1 with and without ionisation. We do not show the column density plot for the Region 2 model with ionisation (R2z85I), but likewise on these scales there is minimal difference to the gas column density with and without ionisation. Nevertheless, we do see that \ion{H}{II} gas is surrounding the areas where sinks have formed, and massive stars are ionising the surrounding gas, where we show the \ion{H}{II} fraction overplotted (top right panel, for Region 1). Note that the scale is such that only the highest integrated \ion{H}{II} fractions appear in the colour bar. We find that the ionisation is altering the structure of the gas, but only on small scales so we investigate this further in our Set 2 simulations. (Note for Region 2 there is further gas which is in the interarm region which is not shown in the figure, as Region 2 has further gas entering from the interarm region).

We show the mass of stars formed, and the star formation rate versus time in the Region 1 and 2 Set 1 simulations, with and without ionisation, in Figure~\ref{fig:sflarge}. There are more stars forming in the Region 1 simulation (the total mass of gas in the simulations is similar), as expected as it is the more strongly converging region, and initially at least has a much higher star formation rate. By contrast for Region 2, there is a more gradual increase in star formation. There is only a very slight reduction in the star formation rate (and mass of stars formed) when ionisation is included, which is just evident for Region 2, suggesting that ionisation is not having a large effect on these scales over timescales of 1--2 Myr. However we look at the star formation rate in more detail in our Set 2 simulations, where we can better resolve the effects of ionisation.

Owing both to the difficulty in seeing the effects of ionisation on these scales, and the difficulty of resolving clusters (since sink particles contain typically 10's or more gas particles), we carried out resimulations of smaller sections where we can achieve much higher resolution (Set 2 simulations). The areas used for the Set 2 simulations are centrally located relative to the Set 1 simulations, and are shown in Figure~\ref{fig:collage}.

\section{Results for Set 2}\label{Set2results}
\begin{figure*}
\centerline{\includegraphics[scale=0.65 ]{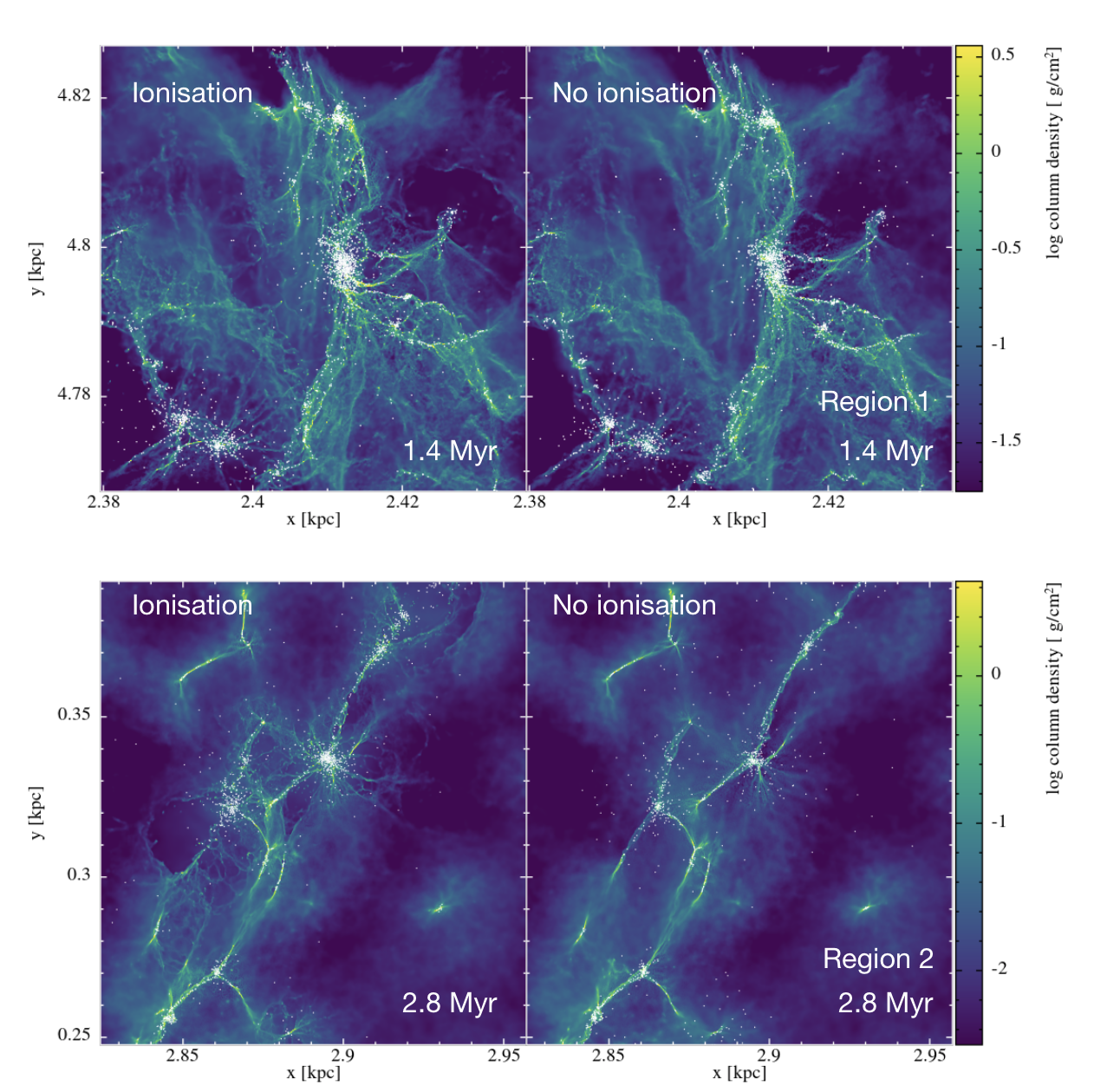}}
\caption{Column density plots are shown for the Set 2 resimulations of Region 1 (top) and Region 2 (lower). Sink particles are overplotted as white dots.}
\label{fig:columndensities}
\end{figure*}

\subsection{Evolution of different regions}
In Figure~\ref{fig:columndensities} we show snapshots of our Set 2 simulations of subregions of Regions 1 and 2, with and without ionisation (R1z2550s, R1z2550sI, R2z2550s and R2z2550sI). As mentioned in Section~\ref{Set1initial}, we take 0 Myr as the start of the Set 1 simulations (198 Myr in the original galaxy simulation), and we note in Region 2 with this reference timescale we do not see any star formation until 0.6 Myr or so later than Region 1, hence the times for Region 2 are later than Region 1. As we will see, the evolution also occurs faster for Region 1, as it is denser, and the clusters merge together.

From Figure~\ref{fig:columndensities}, we see that both regions show one or two larger clusters which have formed, and a number of smaller clusters. The clusters tend to form along filaments, in both cases the filamentary feature spanning from top to bottom of the panels corresponds to the spiral arm, and particularly in the Region 1 simulation (we hereafter refer to the subregions of Region 1 and 2 as simply Region 1 and 2), the most massive looking cluster (as we shall see from Section~\ref{Clusters} this is indeed the most massive cluster) appears at the intersection of filaments. Similar to the Set 1 simulations, there is not a large difference between the large scale structure with and without ionisation. However there are some more detailed differences. In Region 1 (top panels), there is slightly clearer structure in some of the gas features with ionisation. There is also a clear ionisation front in the mid right (from $x\sim2.42,y\sim4.81$ to $x\sim2.435,y\sim4.79$) which is not present in the no ionisation case. The clusters also appear to have small differences, particularly the two large clusters which are in the process of merging in the centre of the panel (we show detailed images of these in Section 4.3.1). The differences are slightly clearer in Region 2, likely in part because this region is less dense. Surrounding the clusters there are clearer dense filaments in the ionisation case, whilst there are also regions which have been substantially effected by ionisation.

\begin{figure}
\centerline{\includegraphics[scale=0.45]{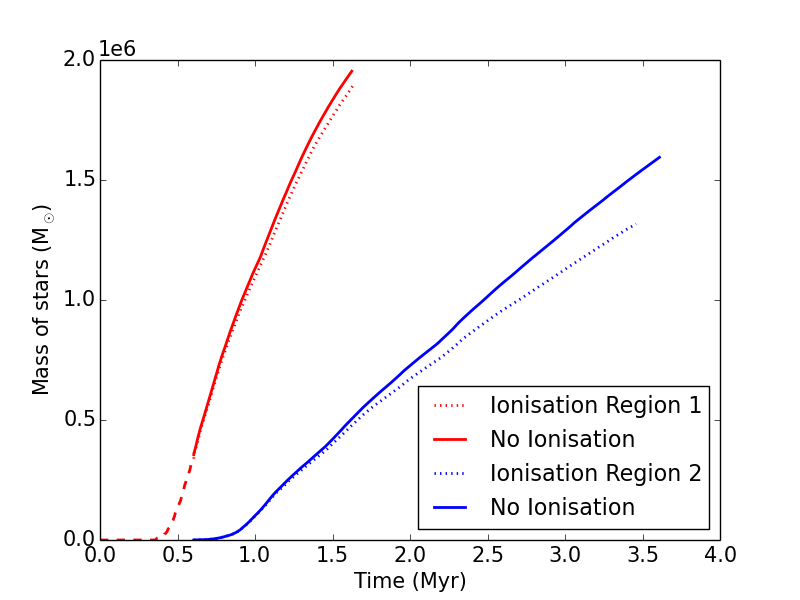}}
\centerline{\includegraphics[scale=0.45]{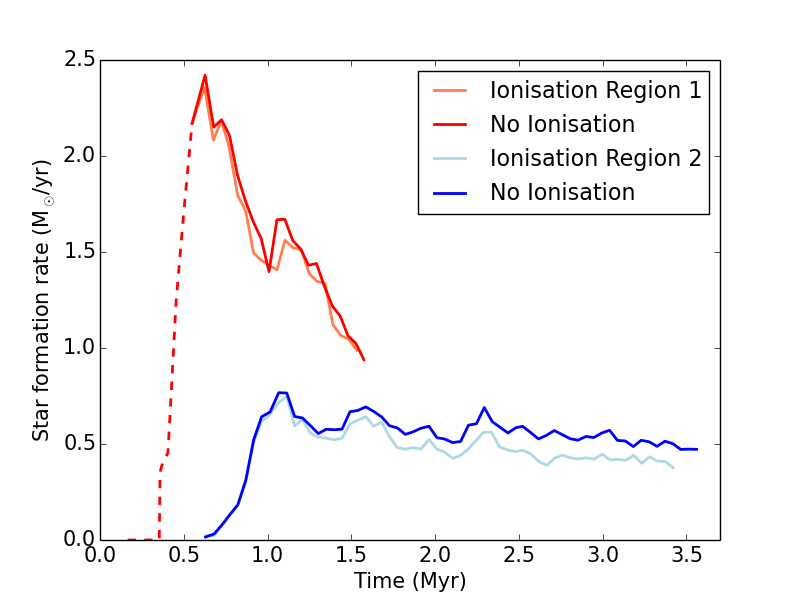}}
\caption{
The mass of stars (top) and star formation rate (lower) is shown versus time for Regions 1 (red) and 2 (blue). The red dashed line shows the star formation rate from the Set 1 simulation, R1z85s. Ionisation has a small impact on the level of star formation for Region 2, for example there is around a 15\% reduction in the stellar mass at the end of the simulation for Region 2 (over about 2 Myr).} 
\label{fig:sfrateregions}
\end{figure}
In Figure ~\ref{fig:sfrateregions} we show the mass of stars fomed and the star formation rate versus time in the two different regions. The mass of stars formed, and the star formation rate, is much higher in Region 1 over an equivalent time, which is not surprising as Region 1 is denser, and has stronger converging flows (Figure~\ref{fig:convergingflows}). In both, the star formation rate appears to be decreasing over time, though the star formation rate is more steady for Region 2. The addition of ionising feedback makes little difference  for Region 1, and only slightly reduces the mass of stars formed. The star formation rate in Region 2 is notably less with ionisation included. The mass of stars is reduced by about 15\% after 2 Myr ($\lesssim$ 1.5 Myr after clusters typically form), or around 20\% after 3 Myr ($\lesssim$ 2 Myr after clusters typically form). Our results suggest there is less reduction in the mass of stars formed in Region 1 compared to Region 2, because it is less dense, although it is difficult to confirm this because Region 1 is not evolved for as long, partly because Region 1 runs more slowly due to the number of sinks and ionising sources, and also because the cluster evolution in Region 1 is more dominated by mergers, and tends towards a small number of massive clusters. We do see however that for an equivalent mass of stars formed, ionisation has had a greater impact on star formation for Region 2 compared to Region~1.

\subsection{Ionisation}

In Figure~\ref{fig:ionxy} we show the ionisation overplotted on the gas column density for Regions 1 and 2. The colour scale shows the rendered \ion{H}{II} fraction, i.e. the \ion{H}{II} fraction integrated through the plane (top panels), and a cross section showing the ionised \ion{H}{II} fraction (lower panels). We also highlight a number of clusters (1,2a,2b,3,4,5 in Region 1; 1,2,3,4 in Region 2) in the top panels which are discussed in Section~\ref{Clusters}. Note that the whole or most of the panel would show a non zero integrated \ion{H}{II} fraction, so our figures only show the regions with the most \ion{H}{II}. For Region 1, there are clearly two regions of more significant \ion{H}{II} gas associated with three of the larger clusters, those labelled Cluster 1, 2a and 2b. The ionisation from Cluster 4 at least partly also joins that from Cluster 1, suggesting that the ionisation regions may be associated with a few distinct clusters.

For Cluster 1, the highest \ion{H}{II} fractions are offset from the cluster itself, as the photoionising radiation preferentially affects the lower density gas. For Region 2, again there are regions of higher \ion{H}{II} fractions in the low density areas close to the large clusters. We show the \ion{H}{II} fraction in the $z-x$ plane in Figure~\ref{fig:ionxz} (Region 1, top; Region 2, lower). This highlights that likewise, the ionised gas is displaced from the clusters to lower densities in the vertical direction as well. The ionised gas above the clusters, at $z\sim0$ is more or less free to escape from the disc. However for Region 1, the ionised gas at $x\sim2.4$, $z\sim -0.02$, is in a more confined region, due to the extent of the gas below the plane. For Region 2, ionised gas simply moves above and below the plane of the disc. The size of the \ion{H}{II} regions in Regions 1 and 2 are around 10--20 pc, whilst observed \ion{H}{II} regions in the Milky Way are typically a few or 10's parsecs \citep{Anderson2014,Tremblin2014,Scaife2013}. We only see the highest \ion{H}{II} fractions close to the clusters at times within timescales of order 0.1 Myr after ionisation commences, whereby the sizes of the \ion{H}{II} regions are comparable with ultracompact \ion{H}{II} regions \citep{Churchwell2002}.

\begin{figure*}
\centerline{\includegraphics[scale=0.6]{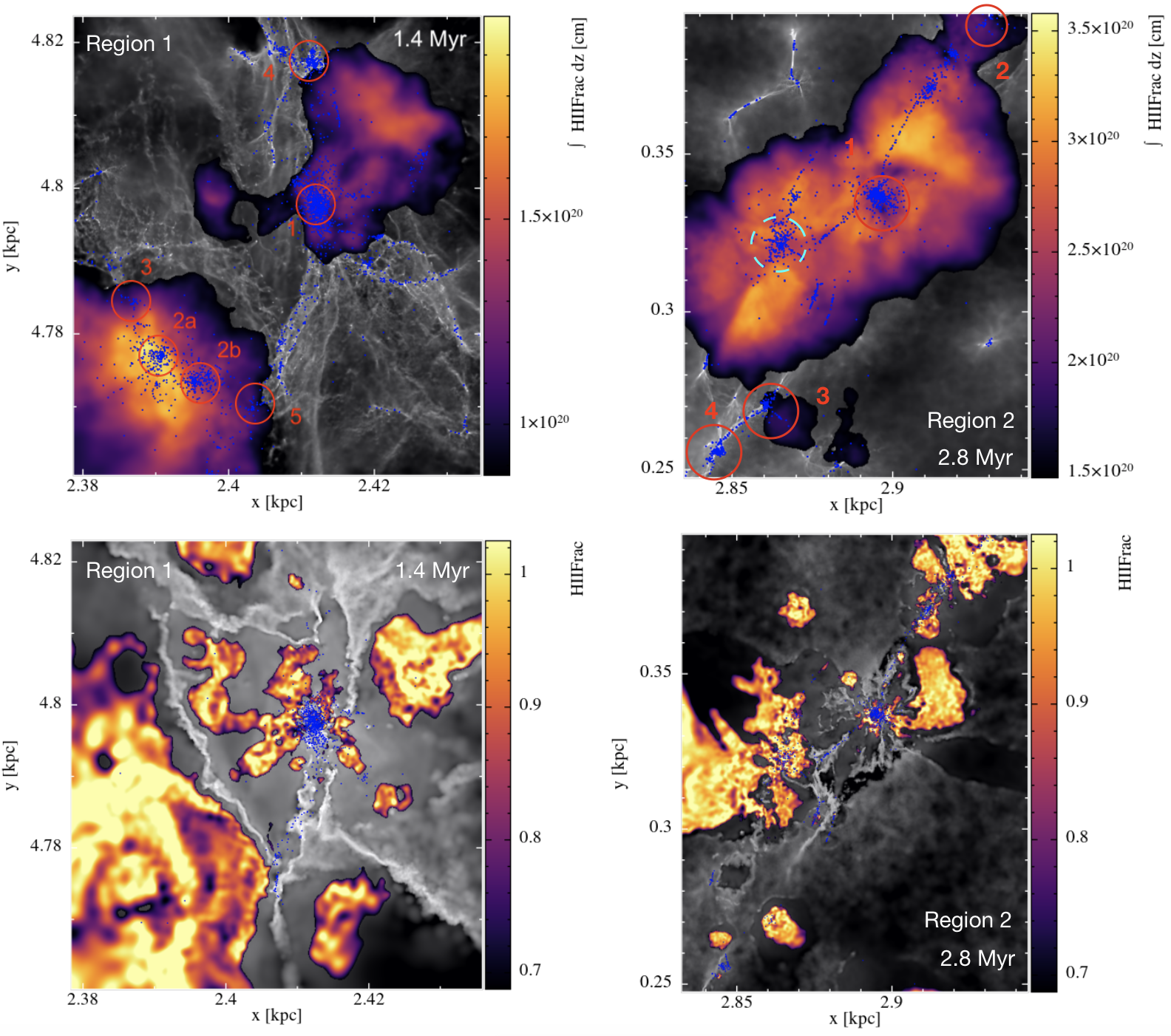}}
\caption{The integrated \ion{H}{II} fraction is overplotted on the column density (shown in grey scale) for 
Region 1 (top left) and 2 (top right) with ionisation. Areas of higher \ion{H}{II} fraction tend to be associated with the most massive clusters with large numbers of ionising sources. Sink particles are plotted in blue, and clusters which are discussed in Section~\ref{Clusters} are highlighted. The lower panels are cross-sections showing the absolute \ion{H}{II} fractions. The cross sections are taken at $z=-0.02$ for Region 1 (lower left), and $z=0$ for Region 2 (lower right).}  
\label{fig:ionxy}
\end{figure*}

\begin{figure}

\centerline{\includegraphics[scale=0.38]{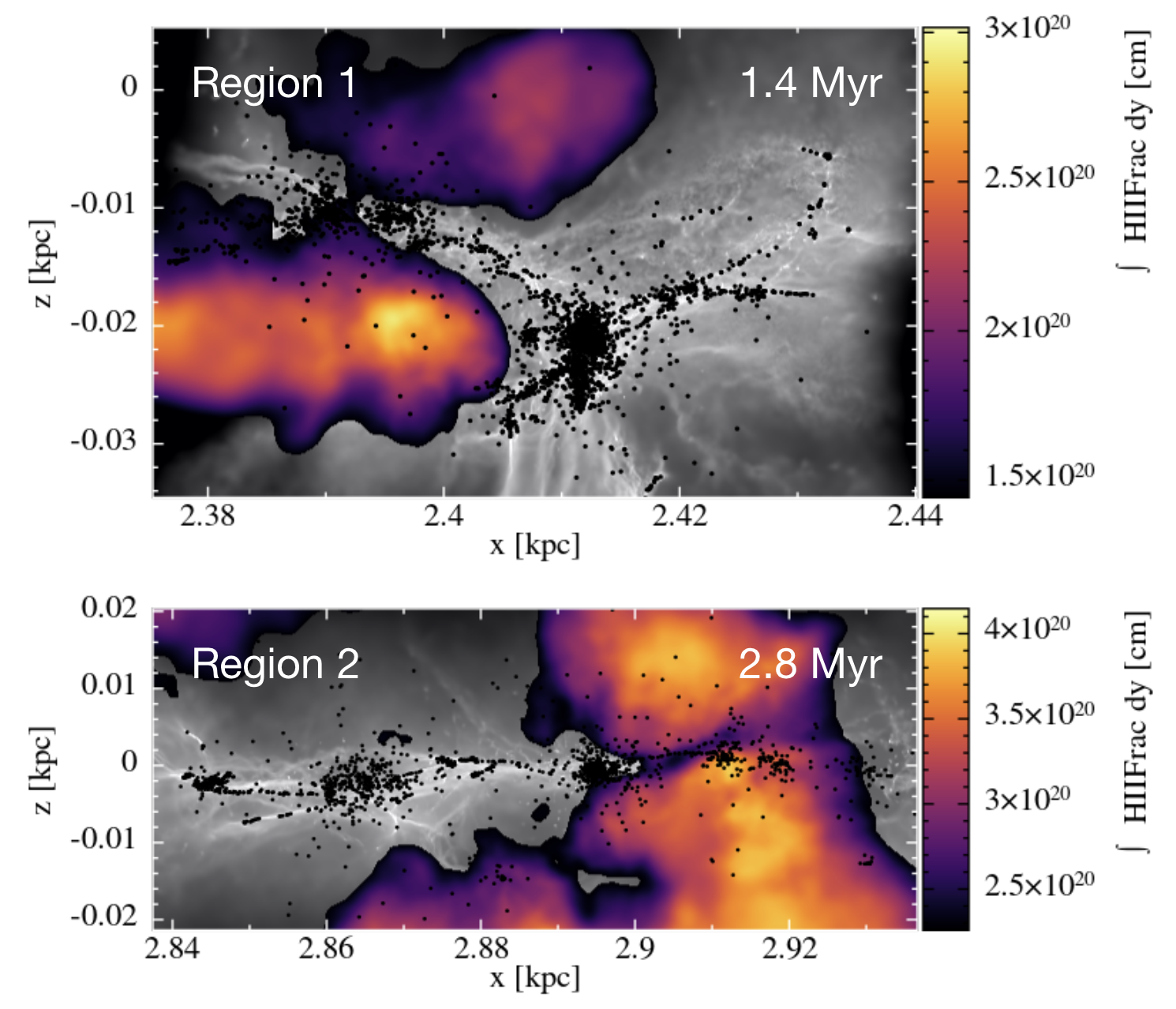}}
\caption{The integrated \ion{H}{II} fraction is overplotted on the column density in the vertical ($z-x$) plane for Regions 1 and 2. The areas with the highest \ion{H}{II} fraction lie above and below the plane.}  
\label{fig:ionxz}
\end{figure}

In Figure~\ref{fig:ionsources}, we show the location of the ionising sinks in Regions 1 and 2 on column density plots with the integrated \ion{H}{II} fraction overlaid. The sinks are coloured according to their flux. The total ionising fluxes of the sinks are $>10^{50}$ (red), $10^{49}-10^{50}$ (green), $<10^{49}$ ergs (yellow), whilst those in blue contain no ionising sources. Sources shown in yellow typically contain one ionising star, those in green one or a few ionising stars and those in red are sinks which typically contain $>10$ ionising stars.  For Region 1 (top panel), there are clearly a large number of ionising sinks associated with the massive cluster at $x\sim2.41$ kpc, $y\sim4.8$ kpc, and perhaps some indication that the most massive are more strongly concentrated at the centre of the cluster. There are also two large clusters with many ionising sources at $x\sim2.39$ kpc, $y\sim4.775$ kpc, which as we see in the next Section, merge together. Both of these regions are also associated with the most intense \ion{H}{II} ionisation. There is also a massive cluster with many ionising sources at the top of the figure ($x\sim2.41$ kpc, $y\sim4.82$ kpc), which is not so strongly associated with \ion{H}{II}, but it is likely that the ionisation predominantly joins the \ion{H}{II} region at $x\sim2.42$ kpc, $y\sim4.81$ kpc.
There are also highly ionising sinks which are not so clearly associated with the very massive clusters, although most are in clusters of other sink particles.

For Region 2 (lower panel), there are two main concentrations of ionising sinks (at $x\sim2.86$ kpc, $y\sim0.32$ kpc and $x\sim2.9$ kpc, $y\sim0.34$ kpc). The most highly ionising sinks are at the centre of these concentrations, and there is lots of \ion{H}{II} gas associated with these regions. Compared with Region 1, although lower in mass, the clusters have evolved for longer (and in relative isolation rather than undergoing mergers), so this may be why the most strongly ionising (and most massive) sources tend to be more at the centre. Other highly ionising sinks are typically associated with other (lower) ionising sinks, often at the nodes of filaments joining together.

\begin{figure}
\centerline{\includegraphics[scale=0.64]{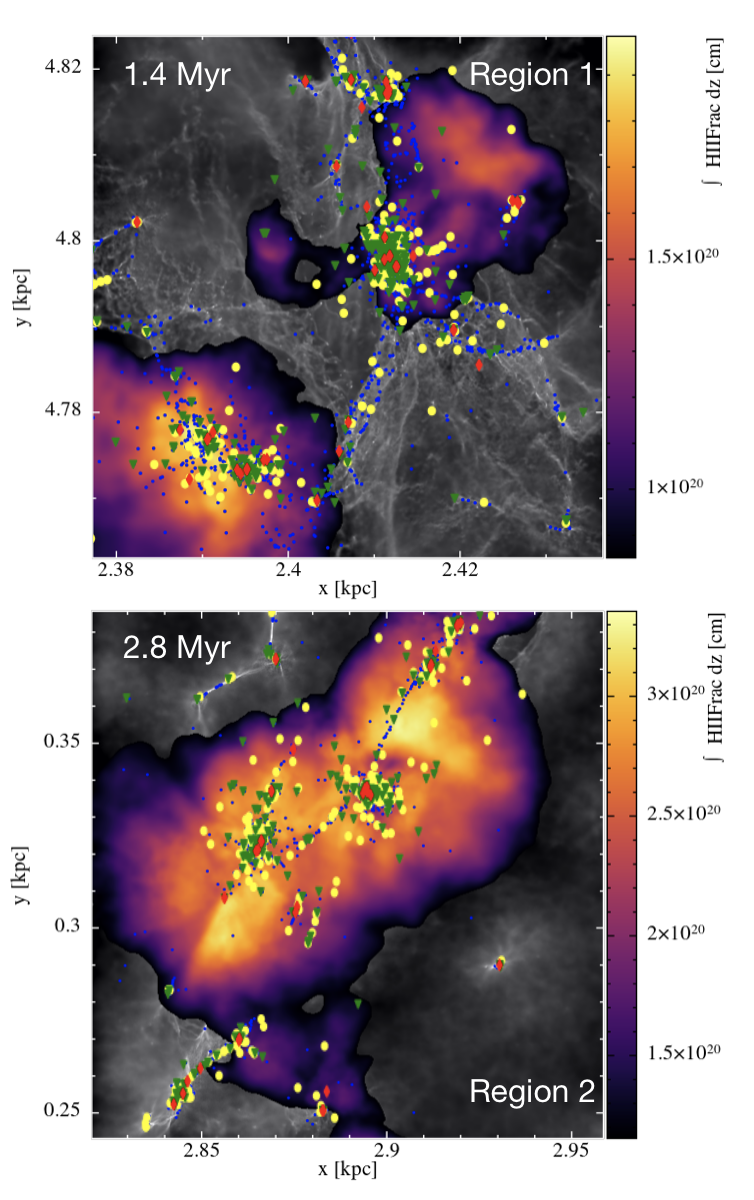}}
\caption{Ionising sources are shown for the simulations of Region 1 (top) and 2 (lower panel). Sources have ionising fluxes of $>10^{50}$ (red), $10^{49}-10^{50}$ (green) and $<10^{49}$ (yellow) ergs are shown. Sink particles which do not contain ionising sources are shown in blue. The integrated \ion{H}{II} fraction is overplotted on the column density. The regions of highest \ion{H}{II} fractions tend to be associated with the largest concentrations of ionising sources which are associated with the most massive clusters.}  
\label{fig:ionsources}
\end{figure}

\subsection{Clusters}\label{Clusters}
In this section we consider the evolution and properties of clusters formed in our Set 2 simulations, which are better able to resolve clusters.
For those in Region 1, some clusters may have started forming before the start of the simulation, i.e. in the Set 1 simulation. Those in Region 2 are all formed in the Set 2 resimulation. Again however we take the timescales for cluster evolution to include the evolution in the Set 1 simulation we well. 

To identify the clusters, we use a friends of friends (FoF) algorithm, similar to our previous work \citep{Dobbs2015,Bending2020}. This algorithm works by grouping together particles which are within a certain distance of each other, which we choose to be 0.5 pc. We show results where we test this distance in the Appendix. We require that clusters contain over 10 sink particles to be selected. We cannot directly convert this to a mass, as the mass of our sink particles varies, but approximately this limits the mass of clusters we can study to $>$1000 M$_{\odot}$. For comparison we also applied the DBSCAN \citep{Ester1996} clustering routine to the same data at one particular time frame. DBSCAN works similarly to the FoF algorithm but requires a minimum number of particles to be within a certain distance. Both algorithms produced very similar results, and in some cases picked out exactly the same set of particles for each cluster. In some cases, the clusters picked out by the FoF algorithm were more extended and asymmetric than the equivalent cluster identified by DBSCAN, which may be an advantage for following clusters whose morphologies are impacted by feedback, though the DBSCAN algorithm may be better at separating clusters which are in the process of merging, or finding spherical clusters. The FoF algorithm identified all the clusters found by DBSCAN, and some additional clusters, but these were all of low mass and of less interest since they are not so well resolved and it is difficult to follow their evolution.

We determine the evolution as follows. For each simulation, we find the clusters at 0.1 Myr intervals. We choose a particular timeframe, for Region 1, we choose 1.4 Myr, for Region 2 we choose 2.8 Myr, and then identify clusters at other timeframes which have the same particles. We focus on clusters which we can pick out over a significant fraction of the simulation, or significant time since they are formed. Note that our results are largely independent of the timeframe chosen provided it is within around half a Myr. For timeframes which are much earlier than this there may have been unbound clusters which have since become too dispersed to be selected according to our our criteria at later time frames. Conversely, some clusters which do not form until late on in the simulations would be missed. This is less relevant for the Region 1 simulations, but more relevant for the Region 2 simulations which run for longer. We also match clusters in our models with ionisation to equivalent clusters in the models without ionisation. This was done by eye, but in all cases it was obvious which clusters were equivalent.

\subsubsection{Cluster evolution: Region 1}
We show in Figure~\ref{fig:ionxy} (top panel) the Region 1 simulation with the integrated \ion{H}{II} fraction over plotted, and the sink particles shown in blue. We highlight with circles a number of clusters (the particles which constitute the clusters are shown in later figures). The evolution of the cluster labelled 1 is discussed below, and those labelled 2a, 2b and 3 are followed in the appendix. The FoF algorithm picked out quite a few other clusters, but some of the smaller ones cannot be followed because the number of particles is small. Furthermore some clusters simply merge with other larger clusters (1, 2a, 2b, 4) so their end masses, and ionising fluxes end up being the same as the larger clusters. Clusters 4 and 5 are other examples of clusters which we could follow the evolution of, and did not merge with larger clusters.
\begin{figure*}
\centerline{\includegraphics[scale=0.5]{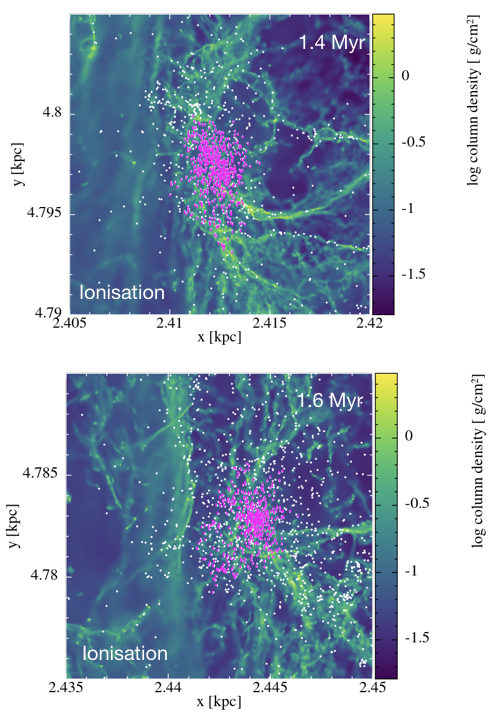}
\includegraphics[scale=0.51]{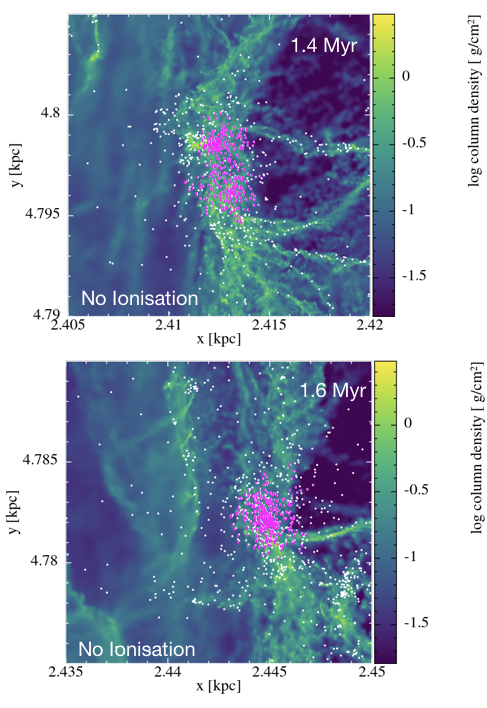}}
\centerline{\includegraphics[scale=0.4]{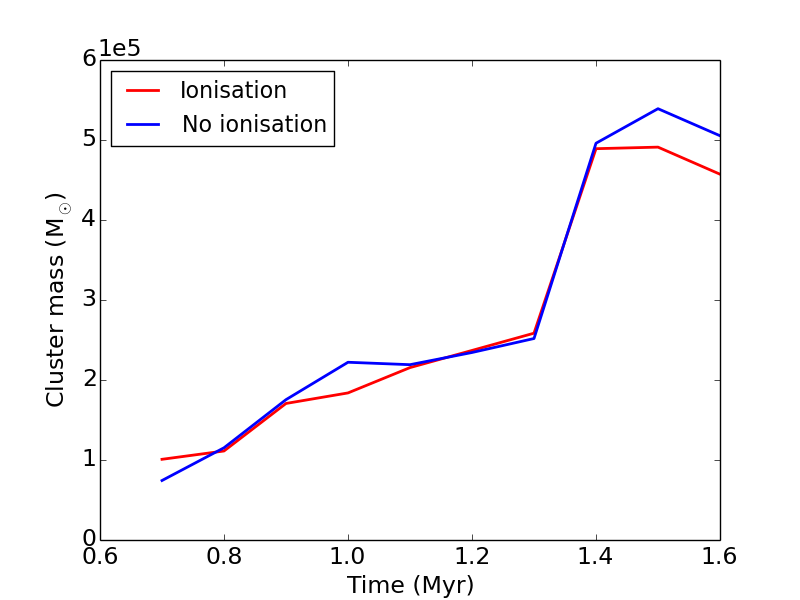}
\includegraphics[scale=0.4]{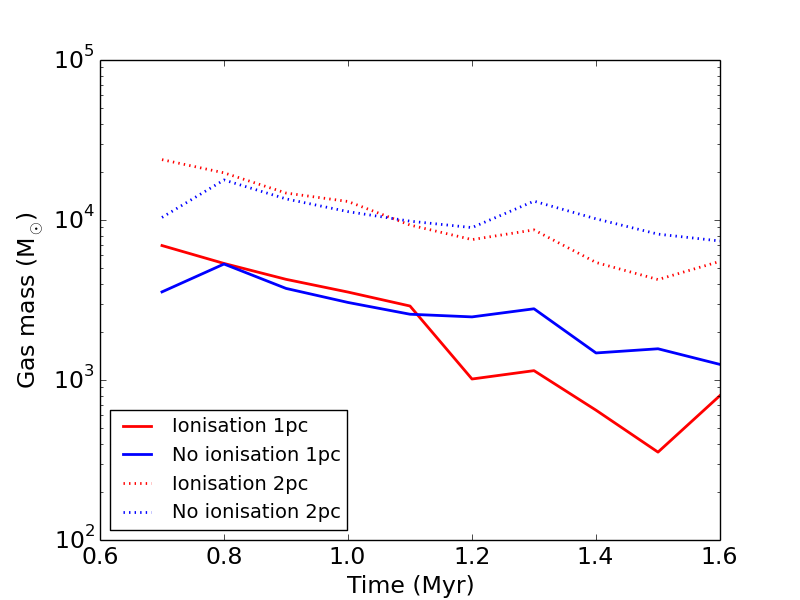}}
\caption{Cluster 1, as shown in Figure~\ref{fig:ionxy} (top panel), from Region 1, is shown a two different times (top panels). The left hand panels are from the simulation without ionisation, the right hand panels with ionisation. In the right hand panels the cluster which is equivalent to that formed in the ionisation simulation is identified by eye. The sink particles which belong to the cluster are shown in magenta. The lower panels show the change in mass of the equivalent clusters with time, and the right hand panel shows the gas mass within radii of 1 and 2 pc of the centre of the cluster.}  
\label{fig:cluster1}
\end{figure*}

In Figure~\ref{fig:cluster1} we show the evolution of Cluster 1 from the Region 1 simulations with (right) and the equivalent cluster without (left) ionisation. We show the cluster at two different times, 1.4 and 1.6 Myr. We also show the evolution of the cluster mass over time, and the evolution of the gas mass associated with the cluster, again with and without ionisation. Cluster 1 forms from the merger of 2 massive clusters (whilst some other clusters also merge to form Cluster 1 they are much smaller). Markedly, we see that in the no ionisation case, the clusters take longer to merge, and the two cores of the clusters are clearly seen at 1.4 Myr in the no ionisation case, whereas with ionisation, they have clearly merged together. We interpret this difference as due to the ionisation removing gas from the cluster region and effectively changing the potential. In our initial runs, with the IMF sampling method in \citet{Bending2020} the ionisation was stronger, the evacuation of the gas very clear, and this effect was even more pronounced. The kinematics of such evacuated gas will be the subject of a future study (in prep).
By a time of 1.6 Myr, the clusters appear to have fully merged in both cases, at least in terms of their morphology they appear to become roughly spherical on very short time scales of $\sim 0.2$ Myr, though we do not consider the dynamics of the clusters.

The lower left panel of Figure~\ref{fig:cluster1} shows the cluster mass versus time. The ionisation appears to have a small effect on the mass of the cluster, reducing the mass from around $5\times10^5$ M$_{\odot}$ to $4.5\times10^5$ M$_{\odot}$, i.e. by around 10\%. The right panel shows the gas mass within radii of 1 pc, and 2 pc from the centre of 
the cluster. We originally looked at finding clouds associated with the clusters. However the region is not that large and quite dense, much of the gas is molecular, so the region could be mostly considered one or two GMCs, or perhaps would be better termed a Giant Molecular Association (GMA). Instead, at each time frame we determine the centre of the cluster, and calculate the gas mass within 1 and 2 pc radii. The effects of the ionisation can be seen at larger radii from the clusters, or equivalently ionising sources, however the effects at these larger distances are more to compress the gas into denser filaments, rather than necessarily effectively disperse the gas. Also, at larger radii of 10 pc or more, that distance is large enough to sometimes include other clusters. Hence we look at quite small distances from the cluster centre. Note also that a radius of 1 pc is quite similar to the typical cluster radii. For both the ionisation, and no ionisation cases, the amount of gas associated with the cluster decreases with time, in the no ionisation case this is due both to additional star formation, and likely the dynamics of the region. However the amount of gas decreases by much more in the case with ionisation, decreasing by 90\% in the case with ionisation versus 65\% without ionisation. The resultant gas mass in the ionised model varies between 25\% and 65\% of that of the no ionisation model in the latter timeframes. There are similar trends when taking a 2 pc radius, though they tend to be slightly less pronounced.

In the appendix we show examples for 2 more clusters from Region 1. Both show a clear decrease in the gas around the clusters with ionisation that is more pronounced than without ionisation. For the clusters labelled 2a and 2b, which merge together, the mass of the cluster after 1.6 Myr is half that in the run with ionisation compared to without ionisation. The gas mass within 1 pc decreases by 80 \% without ionisation and 90 \% with ionisation (though the maximum decrease is 95\%). Cluster 3 is a smaller cluster where the impact of ionisation is clearly visible in the distribution of the stars, however in terms of the mass of the cluster, the ionisation does not have a clear impact. There is a clear effect in the gas though, whereby the gas mass decreases by 40\% in the cluster with no ionisation, but 93\% for the cluster with ionisation. Clusters 4 (figures of these clusters are not shown for brevity) and 5 show little difference in the mass of the cluster (though note that for Cluster 5 the increase in mass of the cluster is minimal) between the ionisation and no ionisation models, but for Cluster 5, the change in gas mass within 1 pc shows an 80 \% reduction for the ionised case, compared to a 64 \% reduction for the non ionised case.

Overall for Region 1, from the 5 clusters, the gas mass decreases in the vicinity of the cluster by 76$\pm$25\% for the ionisation case, compared to 52$\pm$13\% for the non-ionisation case, on timescales of $\sim 1$ Myr. The addition of ionisation appears to limit the mass of the clusters by 0 to 50 \%, again on timescales of $\sim 1$ Myr.
\begin{figure*}
\centerline{\includegraphics[scale=0.47]{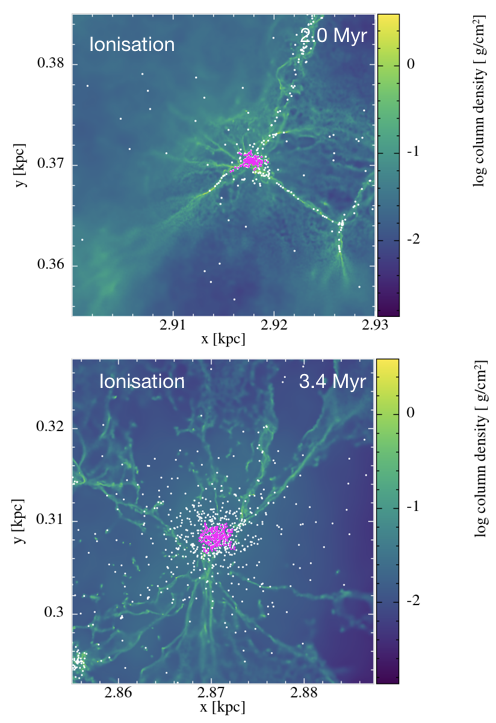}
\includegraphics[scale=0.47]{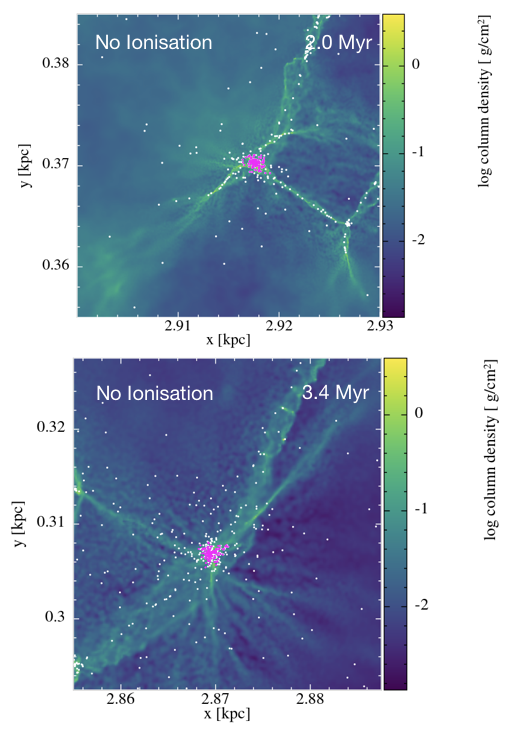}}
\centerline{\includegraphics[scale=0.4]{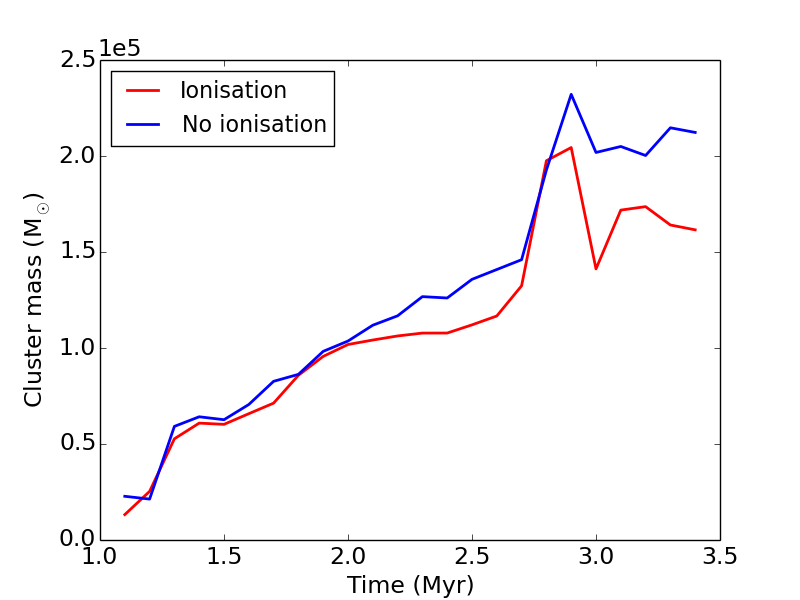}
\includegraphics[scale=0.4]{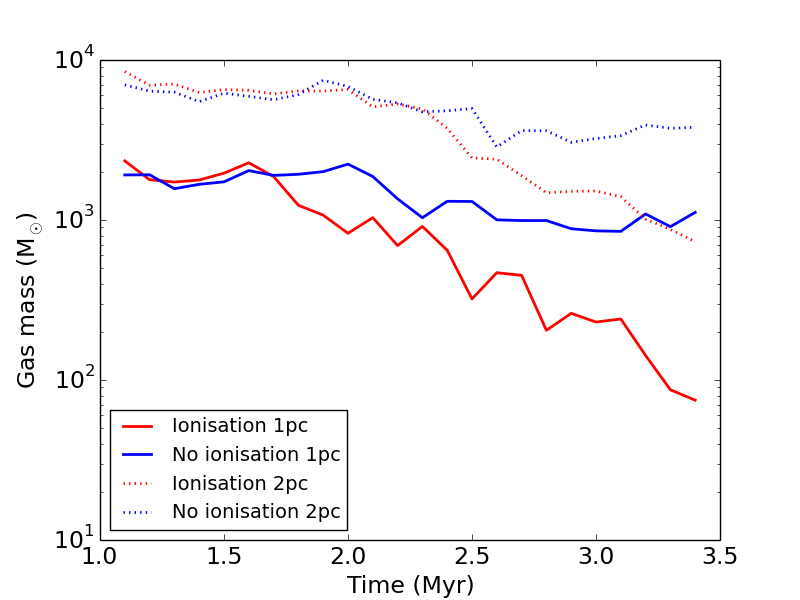}}
\caption{The same as Figure~\ref{fig:cluster1}, but for Cluster 1 highlighted in the Region 2 simulation.}  
\label{fig:cluster2}
\end{figure*}

\subsubsection{Cluster evolution: Region 2}
We evolve Region 2 for longer compared to Region 1, and Region 2 is less dense and has fewer mergers, so we find that it is easier to see the effects of the ionising feedback in this simulation. We show the evolution of the cluster marked 1 on Figure~\ref{fig:ionxy} in Figure~\ref{fig:cluster2}. This is the most massive cluster in this simulation. At the earlier time (top panels) of 2.0 Myr, the equivalent cluster in the simulation with no feedback does not appear so different. However by a time of 3.4 Myr (second panels) the ionising feedback has massively altered the structure of the surrounding gas. The low density gas is smoother due to the ionisation heating up this gas, the gas is visibly lower density right at the centre of the cluster, whilst the filamentary structure around the cluster has the appearance of be being irradiated by the ionising radiation. The mass of the cluster is reduced by 25\% with ionisation, whilst the vast majority of the gas is dispersed from within 1 pc (only 75 M$_{\odot}$, or 7\% of the initial mass remains), whereas 58\% of the gas still remains in the no feedback case. If we instead take gas within larger radii of 2 or 5 pc (the latter not shown), we also see a much greater decrease in mass in the case with photoionisation, compared to the model with no feedback.

The cluster labelled 2 lies at the edge of a region being strongly disrupted by ionising feedback, and again the gas surrounding the cluster is completely different in the case with and without ionisation. This cluster is shown in the appendix. Ionisation again has a strong limiting effect on the mass of the cluster, preventing further growth after around 2 Myr, and decreasing the mass of the cluster by 35\%. Again the amount of gas in the vicinity of the cluster is reduced with ionisation. Interestingly, the low density region to the top right shaped by ionisation is produced by a cluster which we cannot trace at later times, since the stars disperse too much to be detected as a cluster. Clusters 3 and 4 on Figure~\ref{fig:ionxy} show similar behaviour to Clusters 1 and 2, in that all cases show a decrease in the mass of the cluster, and the gas within 1, 2 or 5 pc radius of the centre of the cluster decreases, relative to the no feedback case. The smallest change occurs in Cluster 4, which is perhaps not surprising as this is associated with the least ionisation, and as shown in Figure~\ref{fig:ionsources}, is not associated with any very strongly ionising sources. In this case the cluster mass is only reduced by 5\%. Overall ionisation leads to a reduction in the mass of the clusters by 20$\pm$11 \%, whilst the gas within a 1 pc radius decreases by 
82$\pm$17\% with ionising feedback, and 40$\pm$25\% without feedback, over a period of roughly 2 Myr.

In Figure~\ref{fig:twocluster} we highlight an example for Region 2 where a cluster has had a significant impact on the local gas structure. The cluster highlighted with the cyan circle is associated with very little gas in the ionisation model compared with the model with no ionisation (left panel). In fact, this gas has been pushed out of the cluster by ionisation and now constitutes the filamentary structure 10 pc or so to the right of the cluster. By comparison, without ionisation, the equivalent cluster still occurs along the main filament and still has lots of gas in the vicinity of the sink particles. Unlike the cluster situated to the right of these panels (Cluster 1 in Figure~\ref{fig:ionxy}), this cluster is more or less simply a core of stars at the centre, rather than including a more extended halo (true in the models both with and without ionisation), and it is difficult to follow the cluster over time.

\begin{figure*}
\centerline{\includegraphics[scale=0.35]{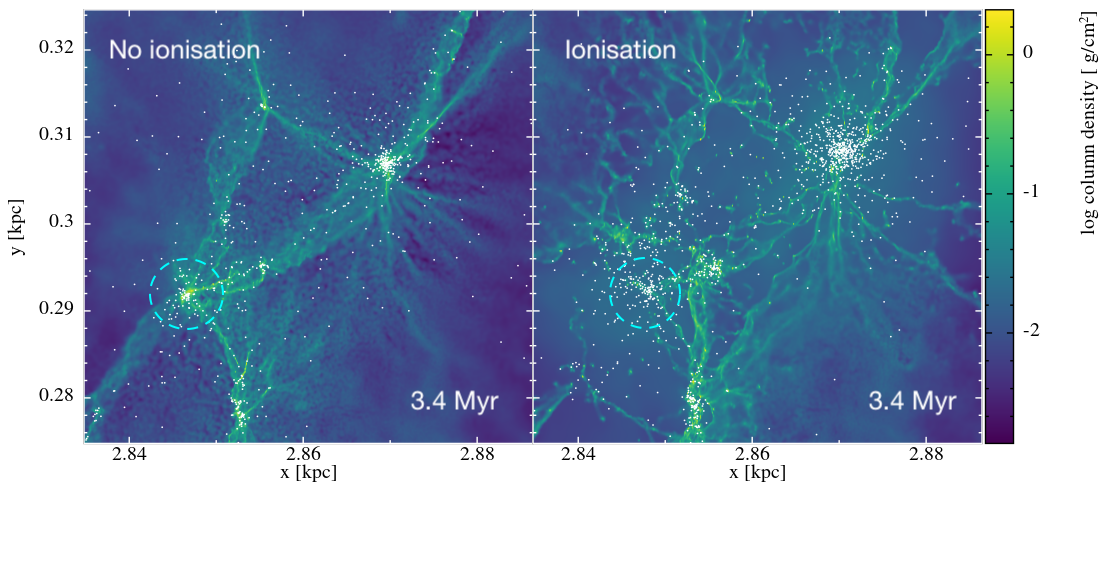}}
\caption{This figure shows the cluster marked by the cyan circle in Figure~\ref{fig:ionxy}, the other cluster apparent is Cluster 1 from Figure~\ref{fig:ionxy} in the models with and without ionisation. Ionisation has a strong effect on the gas for both clusters, but particularly for the cluster indicated by the cyan circle, this cluster is located at the intersection of filaments with no ionisation (left) whereas the filamentary structure is actually shifted by about 10 pc with ionisation (right panel).}  
\label{fig:twocluster}
\end{figure*}

\subsection{Cluster properties}\label{properties}
In this section we consider the overall properties of the clusters in the simulations of Regions 1 and 2 with and without ionisation. We note that results for lower mass clusters may be less reliable since they contain fewer particles, and they may be more transient features (in Region 1 in particular some of these smaller clusters merge with the larger clusters). In Figure~\ref{fig:properties} we show the masses and radii of the clusters (top) as well as observed young massive clusters \citep{PZ2010}, and the ratio of the kinetic to gravitational potential energy (lower panels) for Region 1 (left) and Region 2 (right). We use the full mass, and the maximum radius, rather than the half mass radius, because some of the clusters do not have that many particles. Using the half mass radius would reduce the radii by a factor of between 2 and 3. All results are shown at the end of the simulations, 1.6 Myr for Region 1 and 3.4 Myr for Region 2. Including ionisation has a small effect on the masses of the clusters which form, more so for Region 2, though it is easier to show the effects of feedback by comparing equivalent clusters with and without feedback as we did in the previous section. There is a more marked difference between the clusters found in Regions 1 and 2, with clusters in Region 1 reaching several $10^5$ M$_{\odot}$, and clusters in Region 2 reaching not much more than $10^5$ M$_{\odot}$, and over a significantly longer time period. The most massive clusters form in Region 1 by mergers whereas this does not happen in Region 2. Generally the radii of the clusters are around 1 pc, and do not show much of a trend with mass, similar to observations of clusters in the Milky Way \citep{Pfalzner2013}. The most massive clusters do show larger radii, which could be related to not just their mass but also that they have undergone mergers and not fully settled (see also Rieder et al., submitted.).

The lower panels of Figure~\ref{fig:properties} show the ratio of kinetic to gravitational potential energy for the clusters. The points tend to be scattered around a ratio of $\sim 1/2$, consistent with the clusters being in virial equilibrium.  For Region 2, there is a tendency for the clusters in the run with ionisation to be larger than their counterparts without ionisation, and in most cases the energy ratio is the same or they are less bound.
Any trends seem less clear for Region 1. Although for both regions the clusters appear larger with ionisation, this seems offset by a slightly smaller velocity dispersion which leads to a less obvious difference. There are a few outliers for both regions which are unbound, though these tend to be smaller clusters, and likely clusters substantially effected by feedback as they only occur in the runs with ionisation. Potentially these could more resemble associations compared to YMCs or open clusters.
\begin{figure*}
\centerline{\includegraphics[scale=0.35]{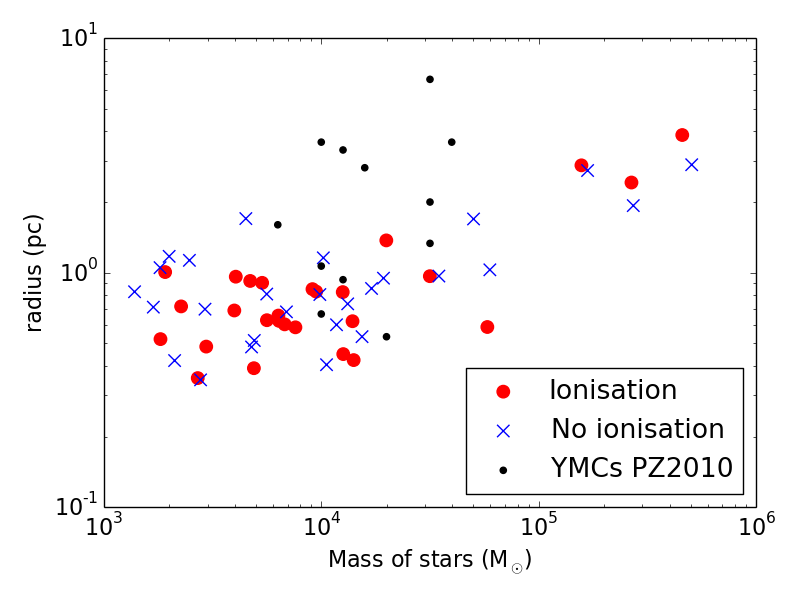}
\includegraphics[scale=0.35]{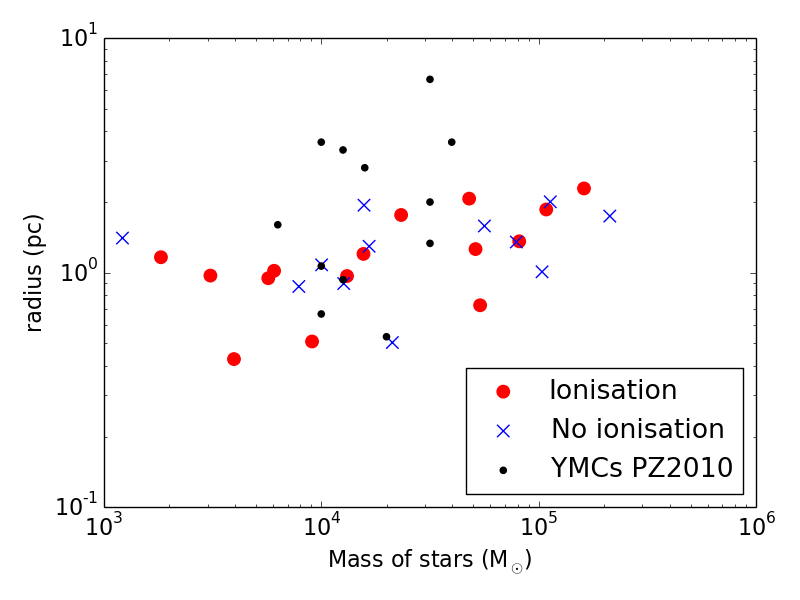}}
\centerline{\includegraphics[scale=0.35]{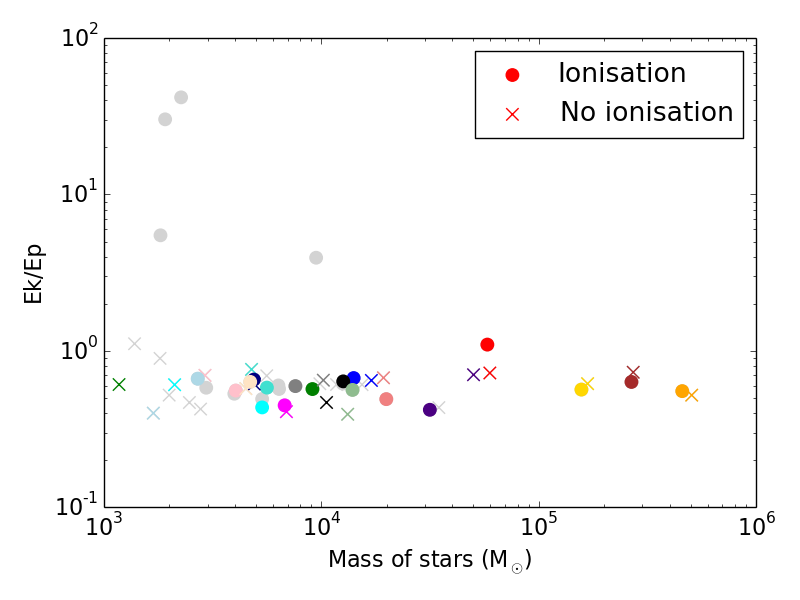}
\includegraphics[scale=0.35]{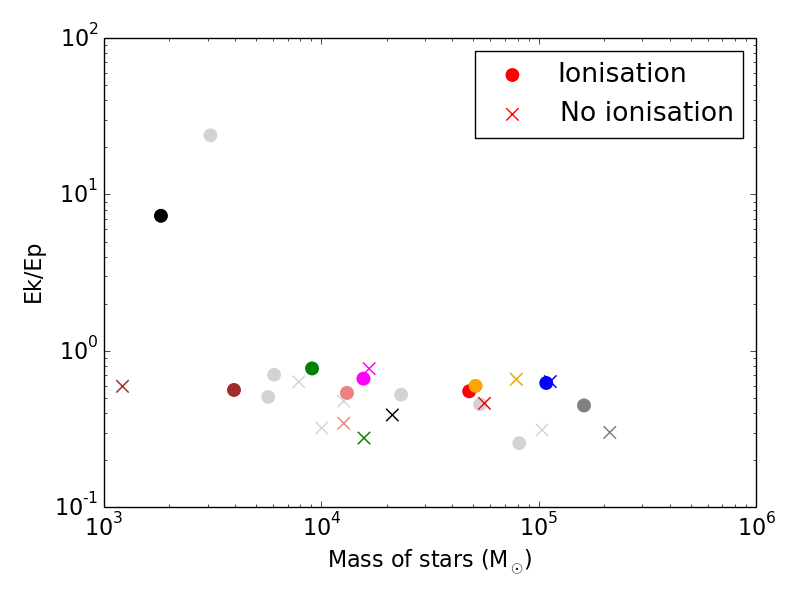}}
\caption{The radius versus mass is shown for clusters found in the Region 1 (top left) and 2 simulations (top right) at a time of 1.6 and 3.4 Myr respectively. Observations of young massive clusters (\citealt{PZ2010} and references therein) are shown as black points, though we caution that there is some dependence of the radius of the simulated clusters both on the FoF algorithm (see Appendix) and the definition of the radius}. The bottom panels show the  ratio of kinetic to gravitational potential energy for the clusters, again for the Region 1 (lower left) and Region 2 (lower right) simulations. In the lower panels, the colours are the same for corresponding clouds in the models. Grey represents unmatched clouds.  
\label{fig:properties}
\end{figure*}

\begin{figure*}
\centerline{\includegraphics[scale=0.35]{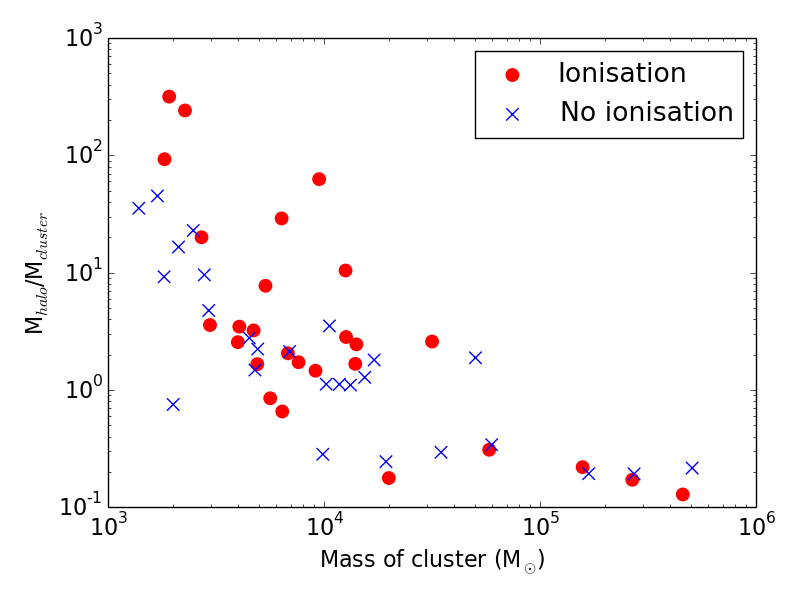}
\includegraphics[scale=0.35]{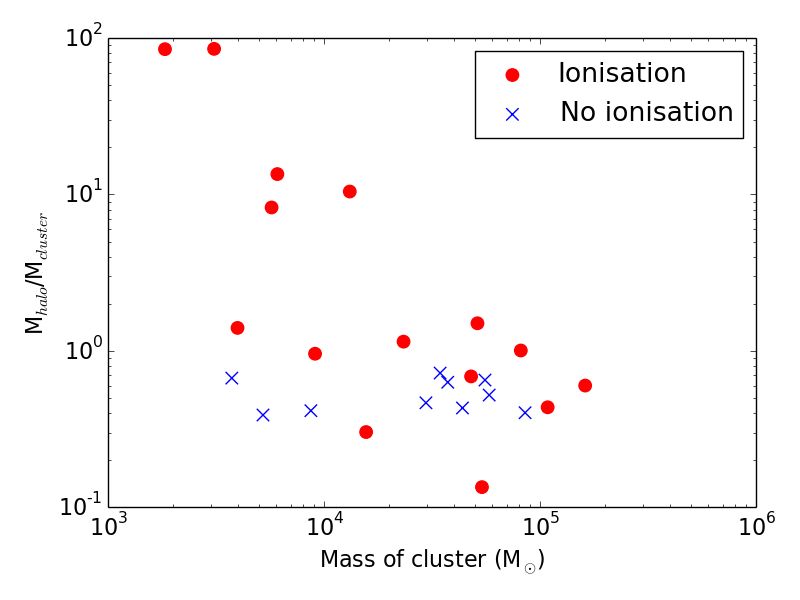}}
\caption{The fraction of the halo mass compared to the cluster mass for Region 1 (left) at a time of 1.6 yr, and Region 2 (right) at a time of 3.4 Myr. The halo mass is defined as the mass of stars within 5 pc (Region 1) and 10 pc (Region 2) of the centre of mass of the clusters excluding the mass of the clusters themselves.}  
\label{fig:halos}
\end{figure*}

Finally we look at the halo of stars surrounding the clusters in our simulations. Our friends of friends algorithm picks out the denser central core of the cluster, but  clusters are generally expected to be surrounded by a lower density stellar halo. Recent observations have found halos, or coronae out to 100 pc or more \citep{Meingast2021}. By eye (e.g. the lower panels of Figures~\ref{fig:cluster1} and \ref{fig:cluster2}), our clusters from the models with ionisation appear to be surrounded by more sink particles (although we note that like \citet{Bending2020}, ionisation tends to produce more lower mass sinks). We estimate the halo stellar masses by taking sink particles which lie within radii of 5 and 10 pc of the centre of mass of a cluster but which are not identified as members of that cluster. In Figure~\ref{fig:halos}, we show the fraction of stellar mass in halos of size 5 pc for Region 1 (left panel), and radius 10 pc for Region 2 (right panel), at times of 1.6 and 3.4 Myr respectively.

For Region 1, Figure~\ref{fig:halos} shows that the fraction of stars contained in the halo is higher with ionisation for the lower mass clusters, but not very different for higher mass clusters. We chose a radius of 5 pc based on the relatively close separation of clusters - if we take a radius of 10 pc instead, the results are similar but with a slightly less discernible trend. We also tried taking the mass within the half mass radius, and taking the halo mass as the mass between this radius and multiples of the half mass radius, and find similar results to those presented. We keep the maximum radius however again simply to ensure sensible particle numbers in our analysis. For Region 2, where the clusters have had longer to evolve and clusters tend to evolve in isolation, we see a more noticeable difference, with the fraction of stars in the halo tending to be higher when photoionisation is included over all cluster masses. We took a radius of 10 pc here, because taking 5 pc gives some very small halo masses, though the trends are the same. The results for Region 2 suggest that gas dispersal occuring in the vicinity of the clusters due to ionisation, and the subsequent stellar motions, leads to greater expansion of the clusters (see also e.g. \citealt{Hills1980,Mathieu1983,Lada1984,Geyer2001,Boily2003, Moeckel2010} where simple prescriptions for gas dispersal are applied to clusters). We also see this from the radii of the clusters, as shown in Figure~\ref{fig:properties} (top right panel).

\section{Discussion and Conclusions}\label{conclusions}
We have performed simulations investigating cluster formation in two different regions along a spiral arm, including photoionising feedback. The dynamics and morphology of the spiral arm are taken from a global simulation of the Milky Way with a live stellar disc. One region (Region 1) has strongly converging gas flows, and appears to be influenced by the merger or collision of two spiral arms. The other region (Region 2) appears to be only a moderately converging region.

Our simulations show that massive clusters form rapidly in converging regions on timescales of Myrs, in agreement with previous work \citep{Dobbs2020,Liow2020}. However whilst these previous simulations only considered simplistic cases of two converging flows (or colliding clouds), we have showed that these are valid in a galactic context. For Region 1, we find that clusters of mass $>10^{5}$ M$_{\odot}$ can form on timescales of order 1 Myr. Region 1 represents the singular most extreme part of our galaxy model, in terms of density\footnote{the criteria for selecting the region was based on divergence but Region 1 is also particularly dense.} and velocities of the gas, so this is where we expect to form massive clusters. These massive clusters form at the hubs of where large scale filamentary structures join together, as observed for the W49A starburst in the Milky Way \citep{GM2013}, and also analgous to massive star formation on smaller scales within molecular clouds \citep{Myers2009,Peretto2013,Baug2018,Kumar2020,Liu2021,Anderson2021}.
For Region 2, we also see massive clusters forming, but the clusters are comparatively less massive and form instead over longer timescales of around 3 Myr. Although dense, Region 2 is not as exceptional as Region 1, and is comparable in properties to numerous massive higher density GMCs in the Milky Way \citep{Rice2016,MD2017,Lada2020}. 

We calculate the cluster formation efficiency, $\Gamma$ \citep{Bastian2008} (see also review by \citealt{Adamo2020}) according to Eqn. 1 of \citet{Johnson2016}. $\Gamma$ is $0.6-0.7$ for Region 1, and $0.4-0.5$ for Region 2. We find that $\Gamma$ decreases over time but there is no clear trend with and without photoionisation. Our values are substantially higher than typically observed, although more comparable with values for highly star forming regions \citep{Johnson2016, Kruijssen2012}. However our clusters are very young compared to the observed clusters. We also note that older clusters may appear more distributed, whilst stars may be ejected from clusters over time, so measuring $\Gamma$ accurately may be difficult and likely only gives a lower limit.

In both Regions, photoionisation does not prevent the formation of massive clusters, although ionisation can be seen to reduce the mass of clusters which form.  Masses of clusters are decreased by up to 50\% but more typically 20\% over timescales of a few Myr with ionising feedback. Ionising feedback can remove nearly all the gas from the vicinity of the cluster in timescales of a few Myr (see also \citealt{Dinnbier2020}, but in practice we see a large spread in both the effect of feedback on the surrounding gas, and resultant cluster masses. We also see that ionisation tends to lead to  clusters which are slightly more extended, and increases the fraction of stars in the stellar halos of the clusters. Ionising feedback visibly alters the morphology of the gas surrounding clusters with strongly ionising sources. Visually the effects of ionisation are similar to individual cloud simulations, with the generation of increasing filamentary structure (e.g. \citealt{Ali2019}). Ionised gas is also preferentially located above and below the plane of the disc and in low density regions, hence ionisation does not always have such a strong impact on cluster formation. Overall, the initial conditions seem to have a stronger role on cluster formation compared to feedback.

The effects of photoionisation may be diminished in our simulations also because we consider particularly dense regions. In relation to this we find that ionising feedback has greater impact in Region 2, in terms of the star formation rate, and in Figure~\ref{fig:appendix3} we can see regions which are highly disrupted by feedback. Previous work has found that the effects of ionisation are less in higher surface density clouds \citep{Ali2019,Hajime2021,Dinnbier2020}. We also see that the efficiency of cluster formation, in terms of the mass of the clusters formed in the presence of ionising feedback compared to no feedback, is still quite high, comparable again to previous results for denser clouds \citep{Grudic2018,Hajime2021}. However we do see an indication that in some cases, ionisation is limiting the masses of resultant clusters (see also \citealt{Hajime2021}). This is in contrast to the work of \citet{Tsang2018} which included solely radiation pressure and found that cluster masses continued to grow. However we have different initial conditions and look at the surrounding gas mass, rather than gas inflow rates so it is difficult to make a direct comparison. We note that even without feedback star formation may be quenched, for example due to the surrounding gas reservoir (which here can change due to the large scale dynamics), so we have directly compared equivalent clusters with and without ionising feedback to take this into account.

We further find that mergers are important for the formation of the most massive clusters. Mergers are much more frequent in Region 1, enabling massive clusters to form on shorter timescales, compared to Region 2, where they are minimal. We see similar findings in Rieder et al., submitted, where again mergers are more common in massive GMCs found in spiral arms with strongly converging flows, and lead to more massive clusters. A similar picture of massive cluster formation by mergers was also put forward by \citet{Howard2019} and \citet{Fujii2015}, although those simulations started with isolated collapsing GMCs. 

In our simulations we have included photoionisation but neglected winds and radiation pressure which we leave to future work. So far our studies (in prep.) indicate that winds do not significantly change the outcomes of our simulations compared to just including ionisation. Although stars formed in our models do not reach ages where they undergo supernovae, we could nevertheless be missing the effects of supernovae from prior generations of stars. We have also assumed that all the gas in sinks is converted to stars, which is likely an overestimate of the star formation efficiency. However even if we assumed only 10\% was converted to stars, we would still see massive clusters form in Region 1 over short timescales, and we would still see the comparative difference between Region 1 and 2. We plan to investigate these factors in upcoming studies.

\section*{Data Availability}

The data underlying this paper will be shared on reasonable request to the corresponding author.

\section*{Acknowledgments}
We thank the referee for a comprehensive report which improved the presentation ad clarity of our paper. Calculations for this paper were performed on the ISCA High Performance Computing Service at the University of Exeter, and used the DiRAC Complexity system, operated by the University of Leicester IT Services, which forms part of the STFC DiRAC HPC Facility (www.dirac.ac.uk ). This equipment is funded by BIS National E-Infrastructure capital grant ST/K000373/1 and  STFC DiRAC Operations grant ST/K0003259/1. DiRAC is part of the National E-Infrastructure. CLD and TJRB acknowledge funding from the European Research Council for the Horizon 2020 ERC consolidator grant project ICYBOB, grant number 818940. Some of the figures in this paper were made using splash \citep{splash2007}.
ARP acknowledges the support of The Japanese Society for the Promotion of Science (JSPS) KAKENHI grant for Early Career Scientists (20K14456).

\appendix
\section{Cluster properties with different friends of friends parameters}
In the analysis presented in the main part of the paper, we used a friends of friends algorithm where we chose a length scale of 0.5 pc as the maximum length scale between sink particles to be able to be part of a cluster. Here we test choosing different length scales, and in Figure~A1 we show properties of clusters where this length scale takes values of 0.25 pc, 0.5 pc and 1 pc. We use clusters from Region 1 at 1.6 Myr, which is the same as the left hand plots of Figure~\ref{fig:properties}. As would be expected the mass and radii of the clusters (left panel) increases with larger distances, though the radii show much greater change than the mass. In the right hand panel we show the ratio of kinetic to gravitational potential energy. This appears robust to the choice of length scale in the friends of friends algorithm, and the clusters appear to mostly be approximately virialised, for different length scales, although the spread appears smaller for the shortest length scale.

\begin{figure*}
\centerline{\includegraphics[scale=0.35]{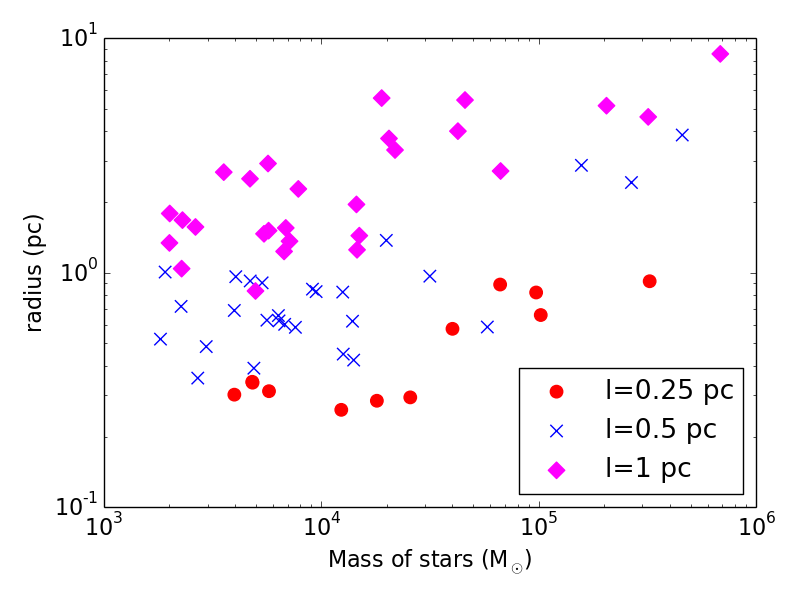}
\includegraphics[scale=0.35]{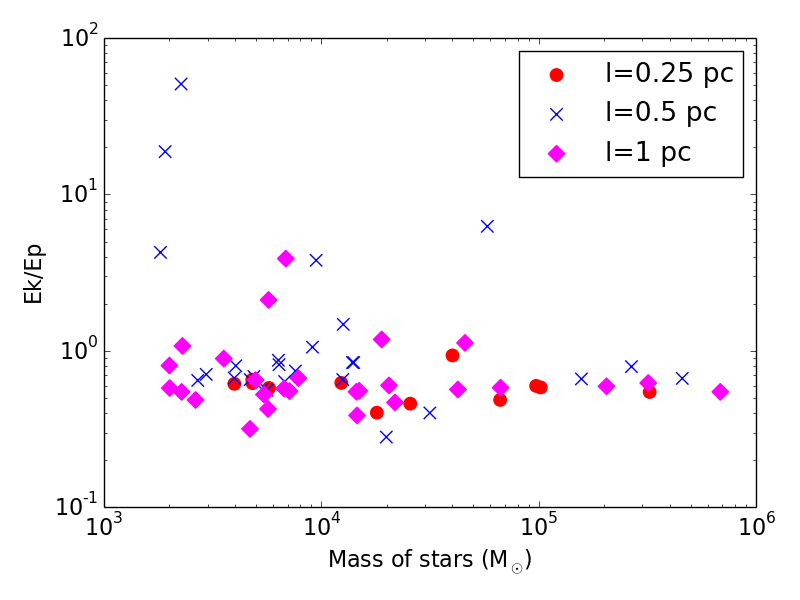}}
\caption{The radius versus mass (left panel) and ratio of kinetic to potential energy (right panel) are shown for clusters found in the Region 1, applying different length scales for our FoF algorithm. The 0.5 pc length scale shows the same results as Figure~\ref{fig:properties}. The clusters are more extended with larger length scales, but approximately virialised over all 3 length scales tested.}  
\label{fig:appendix0}
\end{figure*}

\section{Further examples}
The section shows further examples of cluster evolution. Figures~B1 and B2 are from Region 1, and Figure B3 is from Region 2.
\begin{figure*}
\centerline{\includegraphics[scale=0.47]{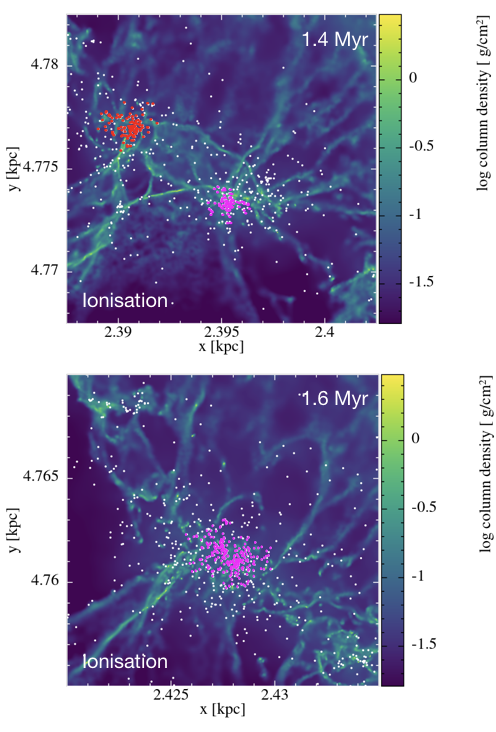}
\includegraphics[scale=0.47]{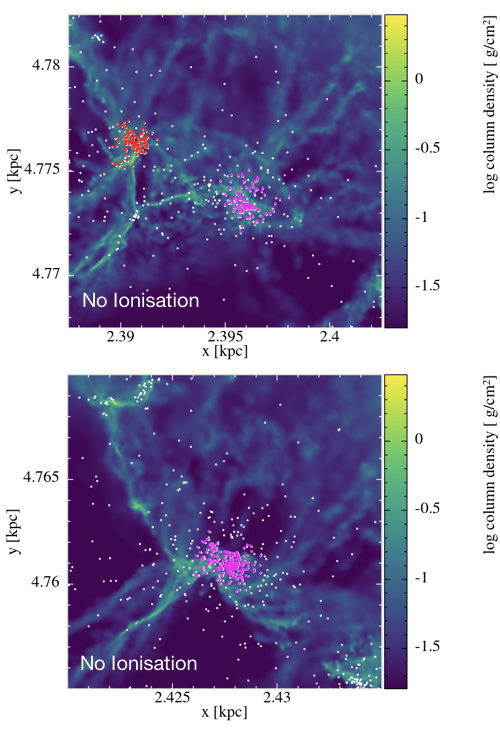}}
\centerline{\includegraphics[scale=0.4]{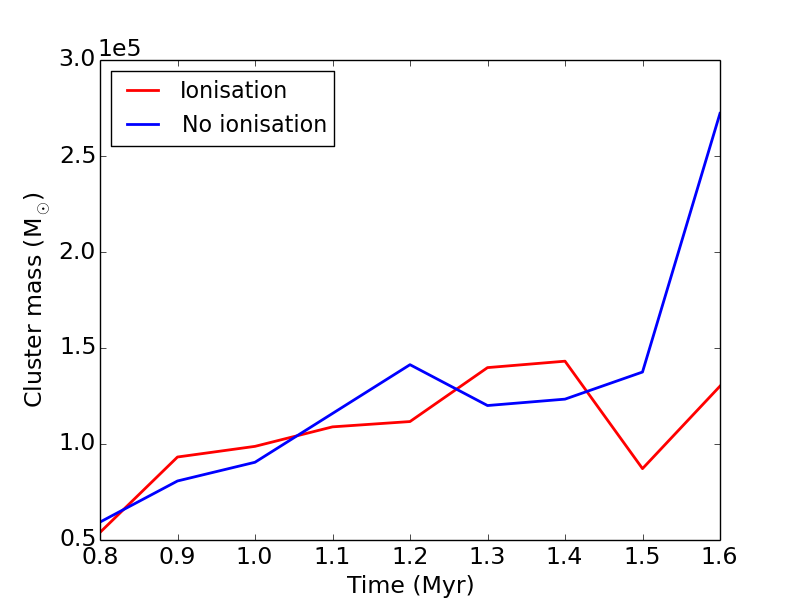}
\includegraphics[scale=0.4]{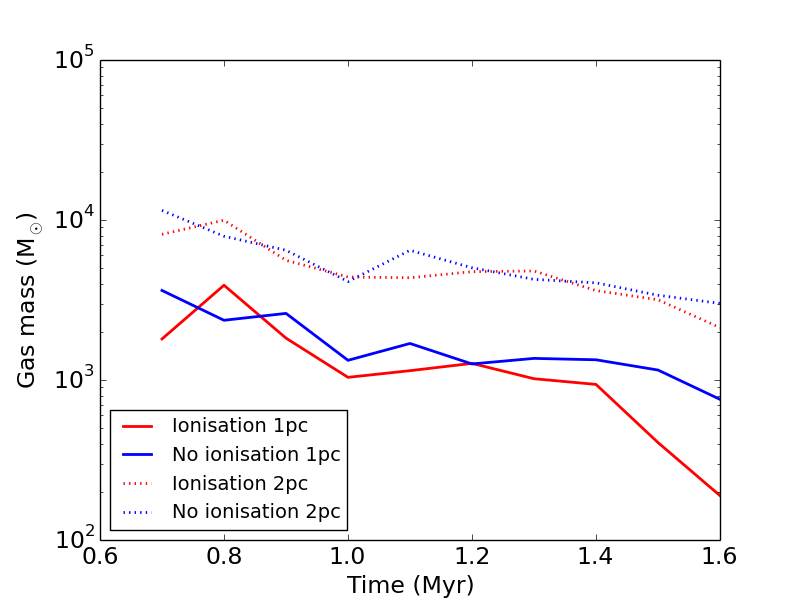}}
\caption{Same as Figure~\ref{fig:cluster1}, but for Clusters 2 and 2b of Region 1.}  
\label{fig:appendix1}
\end{figure*}

\begin{figure*}
\centerline{\includegraphics[scale=0.47]{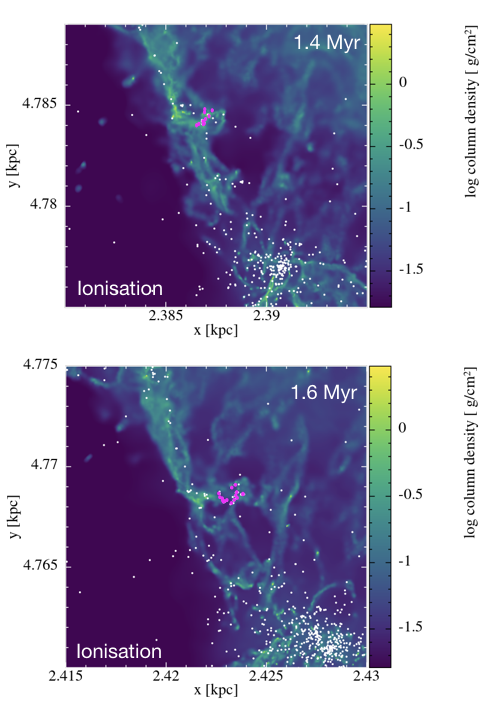}
\includegraphics[scale=0.47]{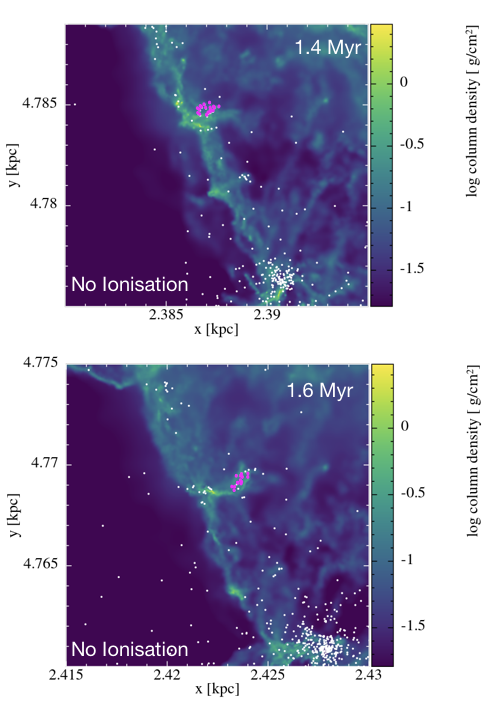}}
\centerline{\includegraphics[scale=0.4]{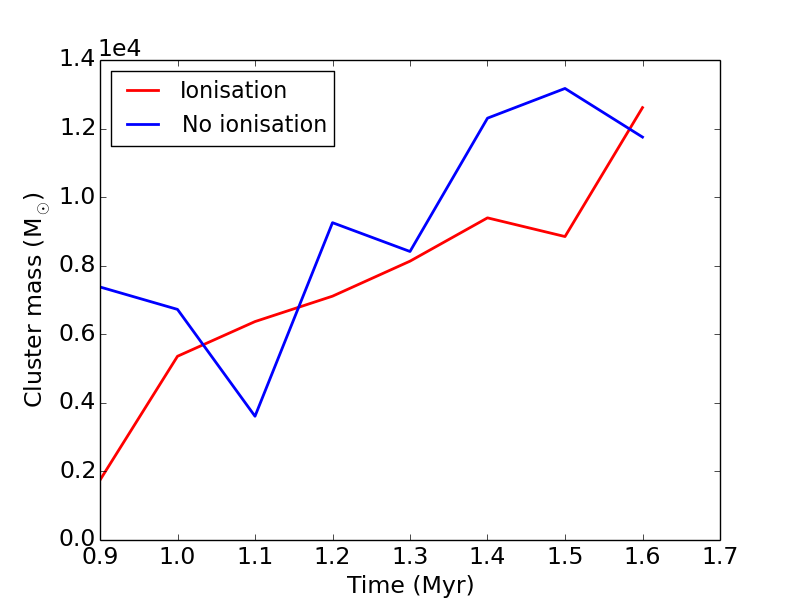}
\includegraphics[scale=0.4]{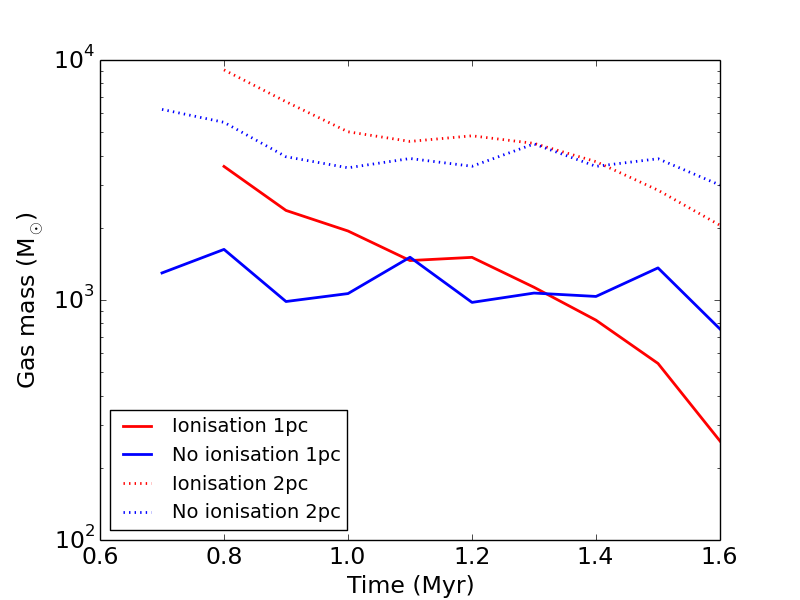}}
\caption{Same as Figure~\ref{fig:cluster1}, but for Cluster 3 of Region 1.}  
\label{fig:appendix2}
\end{figure*}

\begin{figure*}
\centerline{\includegraphics[scale=0.47]{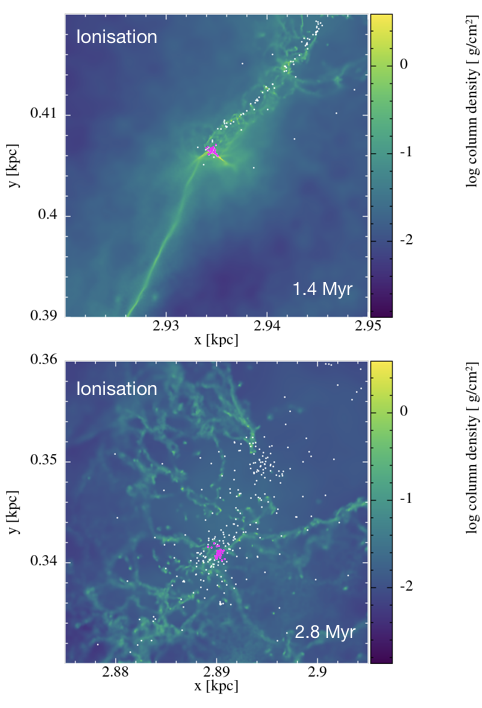}
\includegraphics[scale=0.469]{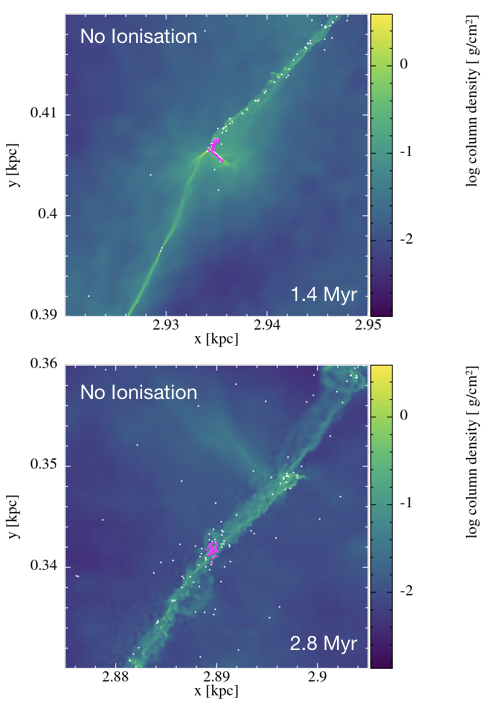}}
\centerline{\includegraphics[scale=0.4]{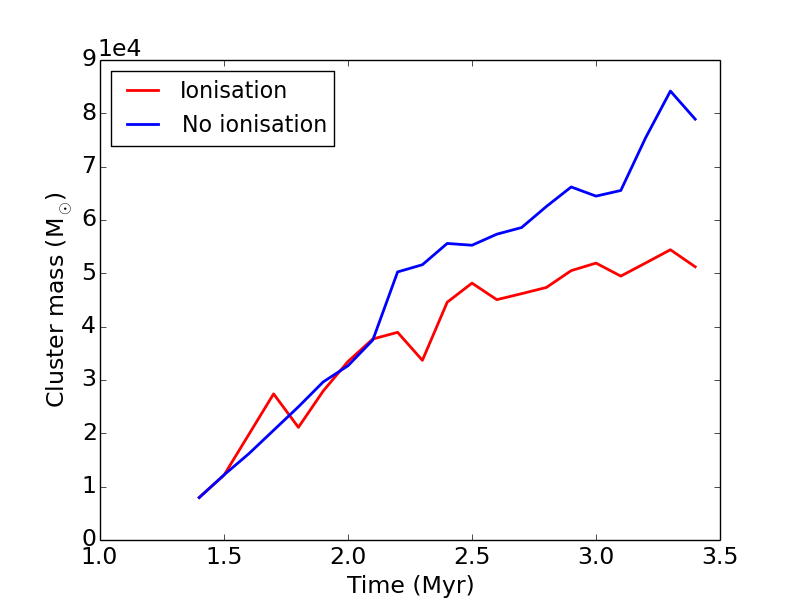}
\includegraphics[scale=0.4]{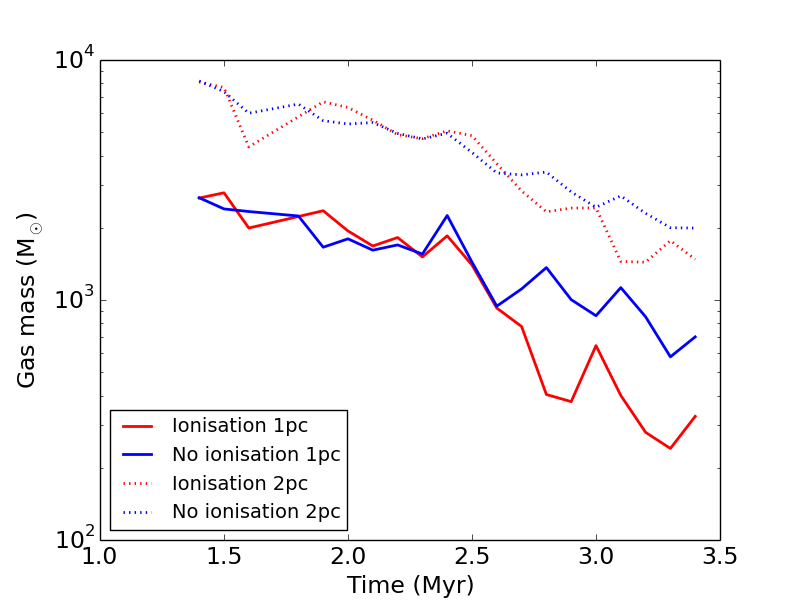}}
\caption{Same as Figure~\ref{fig:cluster1}, but for Cluster 2 of Region 2.}  
\label{fig:appendix3}
\end{figure*}

\bibliographystyle{mn2e}
\bibliography{Dobbs}
\bsp
\label{lastpage}
\end{document}